\documentclass[sigconf,10pt]{acmart}
\newif\ifarxiv
\arxivtrue
\setcopyright{none}
\renewcommand\footnotetextcopyrightpermission[1]{}
\usepackage{tikz}
\usepackage{amsmath}
\usepackage{booktabs}
\usepackage{xspace}
\usepackage{subcaption}
\usepackage{multirow}
\usepackage{xcolor}
\usepackage{enumitem}
\usepackage{makecell}
\usepackage{ifthen}

\usepackage{algorithm}
\usepackage{algpseudocode}

\newcommand\X{VenusRL\xspace}

\begin{document}

\title{\X: A Fully Disaggregated Agentic RL System with Priority Scheduling and Scalable Interaction}

\author{\large
{\rm Mingjun Zhang\textsuperscript{1}},
{\rm Yucheng Li\textsuperscript{2}},
{\rm Menghao Zhang\textsuperscript{2}},
{\rm Shuyong Zhu\textsuperscript{1}},
{\rm Ping Zhang\textsuperscript{3}},
{\rm Xiaohe Hu\textsuperscript{3, 4}},
{\rm Jun Chen\textsuperscript{3}},
{\rm Zhixin Wang\textsuperscript{4}},
{\rm Xutong Wang\textsuperscript{3}},
{\rm He Liu\textsuperscript{3}},
{\rm Yanmin Jia\textsuperscript{3}},
{\rm Shengrong Zhu\textsuperscript{3}},
{\rm Peng Sun\textsuperscript{5}},\\
{\rm Mingjie Zhang\textsuperscript{3}},
{\rm Liming Liu\textsuperscript{4}},
{\rm Jinlong Hou\textsuperscript{4}},
{\rm Yuan Cheng\textsuperscript{4}},
{\rm Yujun Zhang\textsuperscript{1}} \\
\textit{\textsuperscript{1}Institute of Computing Technology, Chinese Academy of Sciences} \quad
\textit{\textsuperscript{2}Beihang University} \quad
\textit{\textsuperscript{3}Infrawaves} \\
\textit{\textsuperscript{4}Shanghai Innovation Institute} \quad
\textit{\textsuperscript{5}Shanghai Qiji Zhifeng Co., Ltd.} \\
}

\begin{abstract}

Agentic Reinforcement Learning (RL) trains LLM agents through multi-turn interactions with external tool environments. 
Its multi-turn nature exposes two systems-level bottlenecks unaddressed by existing agentic RL frameworks. 
First, end-to-end training throughput is constrained by the slowest trajectories to complete, yet optimizing per-GPU utilization alone scatters rollout progress across many groups, delaying the completion of enough groups to unblock the next training step.
Second, tool sandboxes are statically over-provisioned by their declared memory ceilings, leaving most physical memory stranded while replicating near-identical state across sandboxes launched from the same prompt. 
We present \X, a fully disaggregated agentic RL system that addresses both bottlenecks. 
\X's priority-aware action scheduler uses length-prediction heuristics to identify sample groups whose completion is most likely to unblock the next training step, and pushes them ahead of others across batch admission, KV Cache residency, and cross-worker request orchestration. 
\X's environment resource manager combines a memory-aware admission threshold with a template-keyed page-sharing pool, packing more sandboxes per node while preserving strict memory isolation via write-protected page table entry aliasing and copy-on-write. 
Across representative agentic RL workloads, \X achieves up to 4.24× end-to-end training speedup over state-of-the-art baselines and reduces environment cost by up to 89\%.
\end{abstract}

\maketitle
\fancyhead[LO,LE,RO,RE]{}
\pagestyle{empty}

\let\thefootnote\relax\footnotetext{Menghao Zhang and Shuyong Zhu are the corresponding authors.
}

\vspace{-0.2\baselineskip}
\section{Introduction}

Agentic Reinforcement Learning (RL) trains Large Language Models (LLMs) to solve complex tasks through iterative interaction with external environments. 
Driven by this paradigm, the applications of LLMs have evolved from simple language assistants into sophisticated agents capable of solving complex, real-world problems ~\cite{guo2025deepseek, deepresearch2025, jin2025search, lu2025arpo, lu2025deepdive, feng2025group}.
This paradigm also motivates the emergence of industry products such as Claude Code ~\cite{claudecode}, OpenClaw ~\cite{openclaw} and Deep Research ~\cite{deepresearch2025}.
Similar to conventional single-turn RL, agentic RL training is dominated by the rollout stage, which accounts for over 70\% of total wall-clock time ~\cite{qin2025seer, gao2025rollpacker, hu2026taming}.
Existing RL systems mainly optimize this phase by mitigating straggler-induced skewness bubbles ~\cite{fu2025areal, gao2025rollpacker, qin2025seer}.
Beyond this shared characterization, the multi-turn nature of agentic trajectories introduces unique system challenges: an efficient rollout scheduling policy for maximizing end-to-end training throughput, and scalable environment interaction to support such flexible scheduling policy.



First, for rollout scheduling, mainstream RL frameworks ~\cite{slime_github, verl} typically adopt trajectory-level scheduling, where each trajectory is treated as an indivisible execution unit.
This coarse granularity ignores generation--interaction state transitions, thereby leaving GPUs underutilized, which we call \textit{turn-level agentic bubbles}.
Recent works address this problem by decoupling generation controllers from environment controllers, allowing  one trajectory's generation to overlap with another's tool interaction ~\cite{cao2025skyrl, lu2025iirollflash}.
Though this action-level scheduling policy maximizes rollout GPU throughput, it may not accelerate training process since the trainer only waits for enough completed trajectory groups to form the next training batch, while this policy scatters rollout progress across samples.
Moreover, KV Cache pressure and eviction are amplified by multi-turn execution.
A trajectory may first generate, suspend during tool execution, and later resume after its KV Cache has already been evicted by other active trajectories ~\cite{chen2026concur,li2025continuum,kang2026thunderagent}.
This causes repeated recomputation overhead.
What's worse, the action-level scheduling, while effective for eliminating GPU idle time, increases the number of interleaved active trajectories and thereby aggravates this problem.
Both problems pose challenges for an optimized rollout scheduling policy.

Second, for environment interaction, the challenge is not merely the safe execution and isolation of each tool sandboxes, but to sustain the high environment concurrency required by fine-grained rollout scheduling.
Action-level scheduling increases the number of active trajectories so that rollout GPUs can consume ready trajectories while others wait for tool execution. 
However, if the environment side cannot support this concurrency, generation workers are  blocked by interaction-side stalls, preventing action-level scheduling from fully realizing its benefit.
Thus, scalable environment interaction in fact places a strong demand on high-density sandbox deployment.
However, in production, sandboxes are conservatively packed by their configured memory ceilings, even though their actual resident set is far smaller, which leads to "memory stranding" ~\cite{fuerst2022memoryharvesting,li2023pond,javadi2019scavenger}. 
Worse yet, existing systems treat sandboxes as independent instances and overlook the structural property of agentic RL: a sample group consists of multiple rollouts sharing the same prompt. Therefore, trajectories within the same sample group are launched from identical templates and execute nearly identical setup code, so a fraction of their pages are byte-identical. 
This leads to "state duplication" ~\cite{saxena2022memorydeduplication, huang2024trenv} across sandboxes that could otherwise share memory.

From the analysis above, we conclude an efficient agentic RL system requires two complementary capabilities: an effective rollout scheduling policy that maximizes training throughput, and a scalable  environment manager that supports such scheduling policy.
To the end, we propose \X, a fully disaggregated system for agentic RL training.
\X decouples training, rollout generation, and environment interaction into independent components, and introduces two key mechanisms: a \textit{priority-aware action scheduler} for end-to-end training throughput, and an \textit{environment resource manager} for cost-efficient environment interaction.

The priority-aware action scheduler retains the benefits of action-level scheduling by separating generation workers from environment workers.
Additionally, it incorporates priority into its scheduling decisions.
\X defines critical groups as unfinished groups whose completion is most likely to unblock the next trainer batch.
At runtime, \X uses length-prediction-based heuristics to identify such groups and prioritizes their execution.
To enforce this priority, \X co-designs the RL framework and rollout engine to provide priority-aware GPU-slot execution, ensuring high-priority samples are scheduled ahead of others.
Beyond GPU scheduling, \X introduces priority-aware KV Cache residency.
It employs trajectory-aware radix cache and uses a three-tier KV Cache model for admission control and eviction decisions.
This preserves KV Cache entries of high-priority trajectories on a best-effort basis.
Furthermore, \X performs priority-aware request orchestration to mitigate the imbalanced distribution of high-priority samples across rollout workers.
It monitors KV Cache pressure across rollout workers and migrates requests with their reusable prefix cache to reduce KV contention of high-priority samples.

The environment resource manager provides scalable interaction by page-grained and high-density sandbox deployment.
At cluster scope, \X provides scalable interaction by a global controller to allocate sandboxes from a shared environment pool and hide deployment details to the framework.
At node scope, \X introduces page-level sandbox placement and dynamic admission control to maximize deployment density while preventing memory overcommitment.
To exploit similarity among agentic trajectories, \X further builds a group-level page sharing pool.
Trajectories from the same sample group share snapshot pages whenever possible: read-only pages are mapped to the same write-protected physical pages, and private copies are created lazily through copy-on-write upon modification.

We implemented \X in 24k lines of Python, Rust and Golang.
We conduct extensive evaluations to validate the effectiveness of \X.
Experimental data shows that \X outperforms state-of-the-art RL systems, achieving 1.07--3.26$\times$ end-to-end training speedup over Slime ~\cite{slime_github}, 1.06-4.24$\times$ speedup over RollFlash ~\cite{lu2025iirollflash}, and up to 2.67$\times$ speedup compared to ThunderAgent ~\cite{kang2026thunderagent}.
Meanwhile, \X reduces environmental costs by up to 89\% compared with E2B ~\cite{e2bsandbox2024}, a representative and widely-use sandbox platform.

In summary, this paper makes the following contributions:
\begin{itemize}
\item We identify two systems challenges introduced by agentic RL training: multi-turn rollout scheduling on the generation side, and external environment management on the interaction side (\S\ref{2}).

\item We propose \X, a fully disaggregated system for Agentic RL training (\S\ref{3}). \X introduces a priority-aware action scheduler (\S\ref{4}) for maximizing end-to-end training throughput, and an environment resource manager (\S\ref{5}) to provide scalable and cost-efficient environment interaction.

\item We implement \X and evaluate it on representative Agentic RL workloads (\S\ref{sec:evaluation}). Experimental results show that \X achieves up to 4.24$\times$  training throughput improvement and reduces environment resource cost by up to 89\% compared with existing systems.
\end{itemize}

\section{Background and Motivation}
\label{2}

\subsection{Agentic RL}
Agentic RL evolves LLMs into autonomous agents capable of solving complex real-world problems, including AI coding, deep research or embodied AI ~\cite{dou2024stepcoder, qi2024webrl, zheng2025deepresearcher, wang2023voyager, yao2022react}. 
Departing from static and single-turn reasoning tasks ~\cite{chatgpt2022,bai2024digirl,qian2025toolrl}, agents act through dynamic interaction with external environments.
In this paradigm, LLM agents synthesize plans, invoke tools, observe feedback, and refine subsequent actions within a long-horizon execution process. 
A typical workflow of agentic RL training comprises three phases: rollout, reward and training.
In the rollout phase, each prompt is replicated into multiple trajectories, and a rollout worker executes each trajectory by repeatedly alternating between model generation and environment interaction until reaching a terminal state.
Once a trajectory is completed, a reward worker evaluates the quality of agent's actions using a scalar reward signal.
The trainer worker collects finished trajectories groups, estimates the advantage and uses them for policy optimization with algorithms like PPO ~\cite{schulman2017proximal} or GRPO ~\cite{shao2024deepseekmath}.

Agentic RL training can be performed in a synchronous or asynchronous setting. 
Synchronous frameworks ~\cite{hu2025openrlhf, havrilla2023trlx, Sheng_2025} enforce strict on-policy training, where rollout must complete before training begins and all responses generated must use the most recent model weights.
Though this synchronization stabilizes training, the system suffers from severe skewness bubble due to response variance within a single batch ~\cite{gao2025rollpacker, qin2025seer}.
Asynchronous frameworks ~\cite{fu2025areal,slime_github,han2025asyncflow} mitigate the skewness bubble by disaggregating rollout from training.
However, they still fail to address the turn-level agentic bubbles caused by multi-turn interaction.




\subsection{Challenges of agentic RL}
As quantified in previous works ~\cite{qin2025seer, gao2025rollpacker, hu2026taming}, rollout dominates the RL training pipeline, accounting for over 70\% of total wall-clock time.
The rollout phase involves repeatedly transition between GPU-side LLM generation and environ\allowbreak ment-side tool execution.
This iterative loop introduces two unique system challenges: multi-turn rollout pipeline scheduling on the generation side and environment resource management on the interaction side.



\begin{figure}[t!]
    \centering
    \begin{subfigure}[b]{0.48\linewidth}
        \centering
        \includegraphics[width=\linewidth]{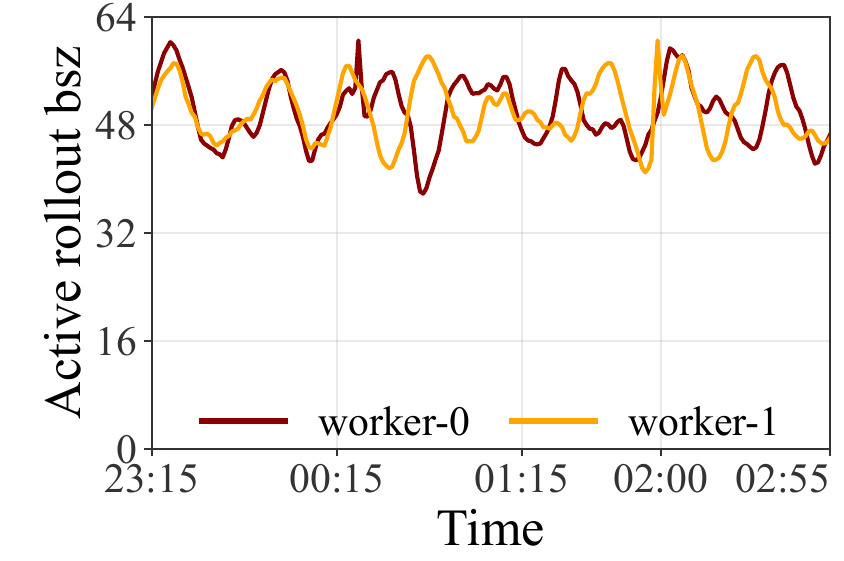}
        \vspace{-1\baselineskip}
        \caption{Active rollout batch size}
        \label{fig:utilization}
    \end{subfigure}
    \hfill
    \begin{subfigure}[b]{0.48\linewidth}
        \centering
        \includegraphics[width=\linewidth]{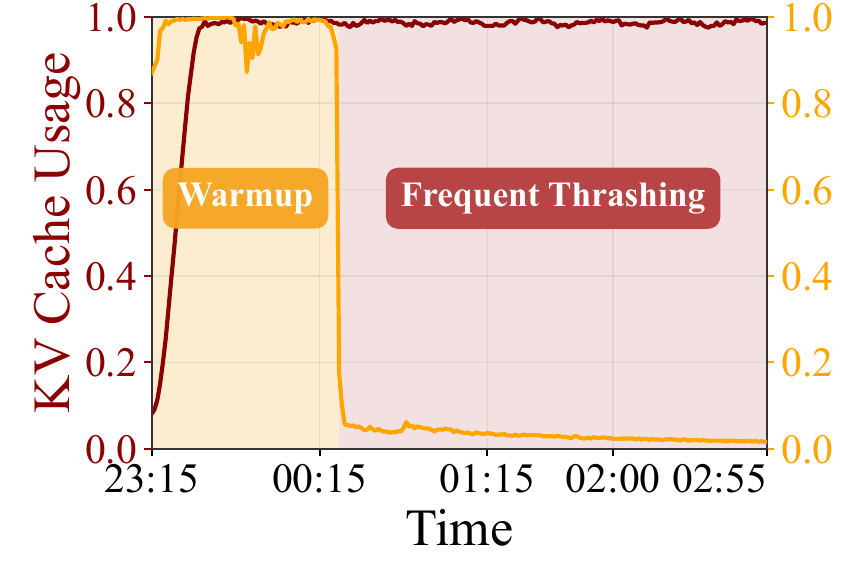}
        \vspace{-1\baselineskip}
        \caption{KV Cache Usage \& hit rate}
        \label{fig:KVCache}
    \end{subfigure}
    \vspace{-1\baselineskip}
    \caption{Profiling results for SWE-agent task (bsz=32).}
    \label{fig:rollout_pipeline_motivation}
    \vspace{-0.7\baselineskip}
\end{figure}

\noindent\textbf{Multi-turn Rollout Scheduling.} 
Agentic trajectories alternate between GPU-side LLM generation and environment-side tool execution.
Mainstream frameworks ~\cite{slime_github,verl} typically schedule rollouts at the trajectory level, and are unaware of the state transitions inside each trajectory.
As a result, when a trajectory pauses for tool execution, it still occupies a generation slot, leaving GPUs underutilized.
We refer to this inefficiency as the turn-level agentic bubble.
We profile the active generation batch size on an online SWE-agent~\cite{yang2024swe} workload.
As shown in Figure ~\ref{fig:utilization}, although the configured generation batch size of each rollout worker is 64, the actual active batch size fluctuates substantially and reaches only  76.5\% on average.
To mitigate this problem, recent works ~\cite{cao2025skyrl,lu2025iirollflash} disaggregate environment controller from generation controller, enabling one tool execution to overlap with the other's generation, which we call action-level scheduling policy.
Though this policy successfully improves rollout GPU utilization, it may not translate into end-to-end training speedup.
Since generation progress is scattered across sample groups, action-level scheduling may even delay this step's training progress (details in \S\ref{sec_problem_formulation_priority}).

Meanwhile, the multi-turn nature of agentic trajectories also introduces persistent KV Cache pressure and unexpected eviction as tokens accumulate across turns. 
Under limited HBM capacity, the system may be trapped in the cycle of eviction and recomputation: each trajectory generates, waits for tool execution, and later resumes after its KV Cache has been evicted by others.
Figure~\ref{fig:KVCache} illustrates this severe KV thrashing.
Although KV Cache usage remains constantly near saturation, the cache hit rate drops sharply and stays low, exposing substantial redundant recomputation.
To mitigate this problem, hierarchical storage backends ~\cite{qin2025mooncake,hicache} extend KV Cache capacity by offloading KV pages to lower-tier storage.
However, they introduce extra reload overhead, which prior work has shown still inefficient ~\cite{chen2026concur}.
Continuum~\cite{li2025continuum} keeps KV Caches alive for a fixed time window using TTL-based management, while ThunderAgent~\cite{kang2026thunderagent} evicts trajectory KV Cache with a shortest-first policy to preserve the continuity of long-context trajectories.
However, these designs are still agnostic to the inner characteristics of agentic RL training. 
\noindent\textbf{Environment Resource Management.}
\begin{figure}[t!]
    \centering
    \begin{minipage}[b]{0.48\linewidth}
        \centering
        \includegraphics[width=\linewidth]{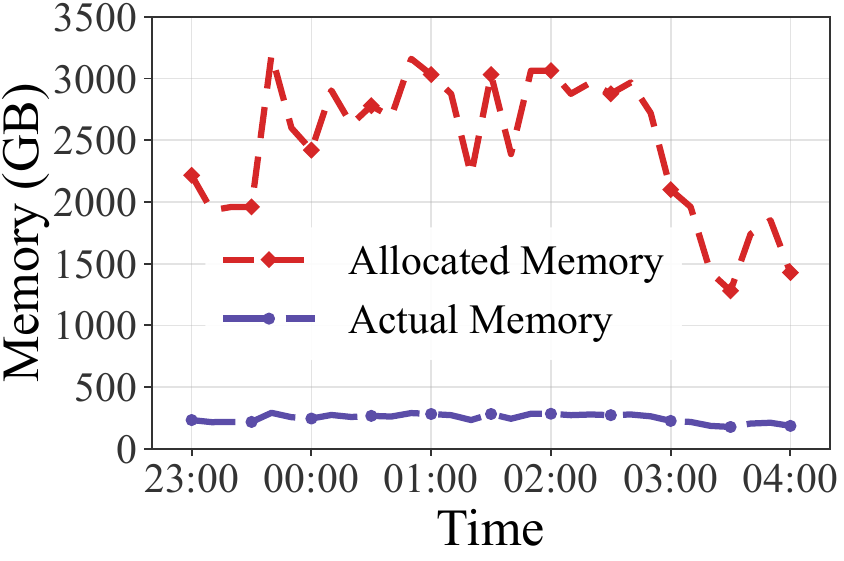}
        \vspace{-2\baselineskip}
        \caption{Online sandbox memory usage.}
        \label{fig:online_sandbox_memory_usage}
    \end{minipage}
    \hfill
    \begin{minipage}[b]{0.48\linewidth}
        \centering
        \includegraphics[width=\linewidth]{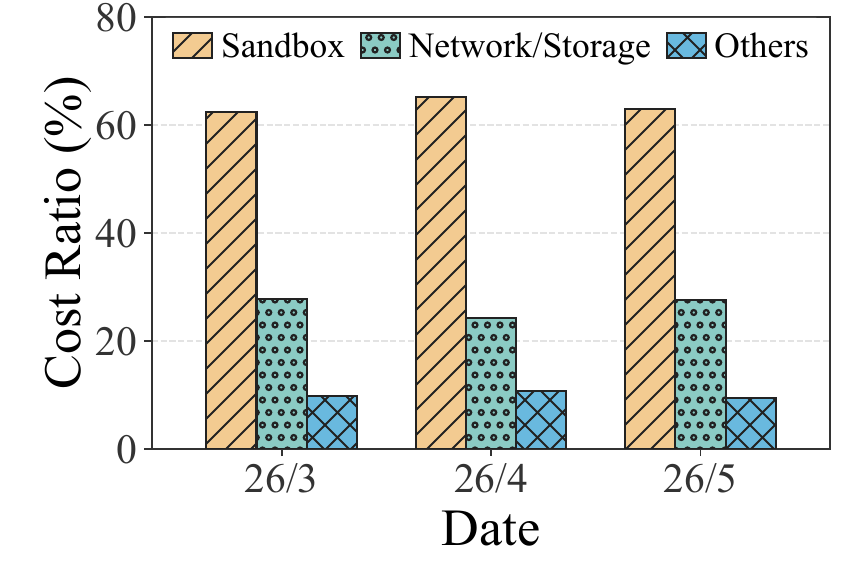}
        \vspace{-2\baselineskip}
        \caption{Non-GPU Agentic workload expenses.}
        \label{fig:agentic_workload_expense}
    \end{minipage}
    \vspace{-0.9\baselineskip}
\end{figure}
Agentic RL relies on external environments for tool interaction.
Such interactions require strong isolation to prevent unsafe tool actions, dependency conflicts, and cross-task interference. 
Existing systems ~\cite{wang2025openhands, yang2024swe, dai2026cuda} therefore commonly place each task in an isolated sandbox.
Modern agentic sandboxes are deployed as microVMs ~\cite{agache2020firecracker, cloudhypervisor} with pre-built images and fixed memory allocations.
In production, memory capacity is the primary bottleneck to sandbox density (Details in Appendix ~\ref{appendix-sandbox-density}).
Unlike CPU and network resources, which can be oversubscribed with graceful performance degradation, sandbox memory is typically provisioned as a hard capacity constraint: once memory is exhausted, no additional sandboxes can be created.
Prior works improve memory deployment density through execution-state reuse and resource elasticity.
TrEnv ~\cite{huang2024trenv} proposes a repurposable sandbox design and facilitates remote memory state sharing by implementing a new Linux namespace API, while ARL-Tangram ~\cite{xiao2026arl} formulates external resource management as an action-level optimization problem and dynamically allocates suitable resources to each sandbox during execution. 

However, none of these works fully address the following two sources of memory inefficiency for agentic sandboxes.
First, to guarantee stability, production deployment therefore limits concurrency by configured peak memory rather than observed usage, introducing "memory stranding" ~\cite{fuerst2022memoryharvesting,li2023pond,javadi2019scavenger}.
Moreover, treating each sandbox as a fully independent instance prevents reuse opportunities of shared runtimes, libraries, and dependencies, resulting in "state duplication"  ~\cite{saxena2022memorydeduplication, huang2024trenv}.
For ARL-Tangram, an additional limitation is that its elasticity relies on lightweight container-level reconfiguration and cannot be efficiently extended to strongly isolated microVM sandboxes, where resources are provisioned at the VM boundary.
Our production profiling confirms this inefficiency: as shown in Figure~\ref{fig:online_sandbox_memory_usage}, sandbox memory usage on a physical node remains only around 8\% on average, while Figure ~\ref{fig:agentic_workload_expense} shows that environment sandboxes contribute more than 60\% of production non-GPU cost over the past three months.

\section{\X Design Overview}
\label{3}


\begin{figure}[t]
     \centering
     \includegraphics[width=1\linewidth]{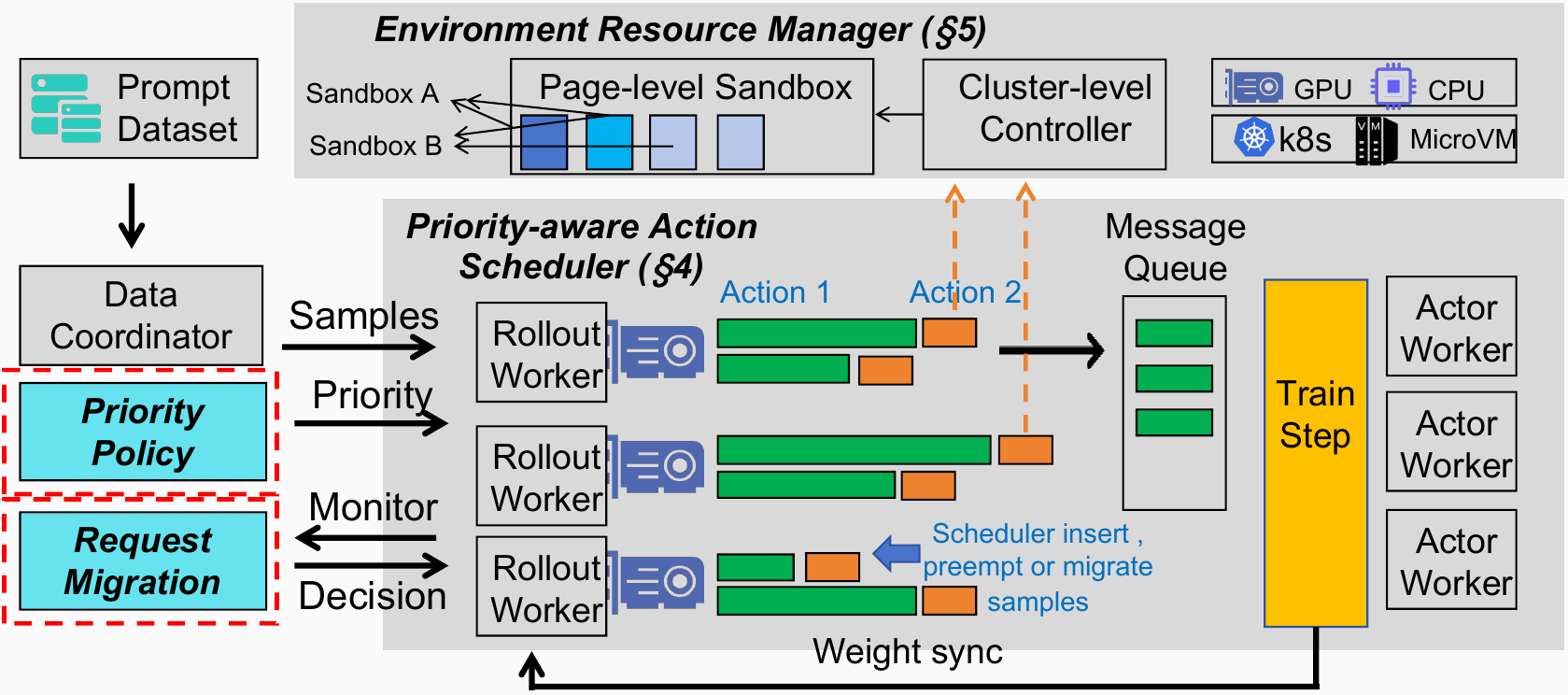}
     \vspace{-1.5\baselineskip}
     \caption[Caption for LOF]{System overview and workflow of \X.}
     \label{fig:overview}
     \vspace{-1\baselineskip}
\end{figure}

To address the challenges above, we propose \X, a fully disaggregated system for agentic RL training, as shown in Figure~\ref{fig:overview}. 
Fundamentally, \X follows the multi-controller paradigm adopted by prior works ~\cite{wang2025distflow, zheng2026astraflow}.
Specifically, the data coordinator serves as the centralized control plane for dataset loading, prompt preprocessing, and sample scheduling. 
Each rollout worker continuously fetches samples from the data coordinator using a dedicated background thread and executes generation asynchronously. 
When a sample group finishes generation, the rollout worker computes its advantages and streams the group to actor workers for immediate gradient computation. 
Once enough sample groups form a global batch, actor workers accumulate all computed gradients, and trigger a synchronized gradient computation and update across all available GPUs.
The updated actor weights are then synchronized back to rollout workers before the next RL iteration.

Unlike previous works ~\cite{zhong2025streamrl,fu2025areal,slime_github} that only decouple rollout from training, \X further separates environment interaction from rollout execution.
Building on this architecture, \X introduces two key components: a priority-aware action scheduler to maximize end-to-end training throughput, and an environment resource manager to provide scalable and cost-efficient environment interaction.

\noindent\textbf{Priority-aware Action Scheduler.}
The priority-aware action scheduler decouples generation workers from environment workers, scheduling samples at action granularity. 
The scheduler incorporates priority into scheduling policy, and uses a lightweight fine-tuned predictor to estimate each sample's remaining generation length at turn boundaries for identifying high-priority sample groups.
Once high-priority groups are identified, the scheduler prioritizes their samples across GPU-slot execution, KV Cache residency and runtime request orchestration.

\noindent\textbf{Environment Resource Manager.}
The environment resource manager provides scalable interaction by page-grained and high-density sandbox deployment.
At cluster scope, the manager maintains a global controller that allocates sandboxes from a shared environment pool, hiding the deployment details from the RL framework.
At node scope, \X introduces page-level sandbox placement and dynamic admission control to maximize node deployment density while preventing memory overcommitment.
Additionally, \X exploits the strong similarity among agentic RL sandboxes through a group-level page pool to reuse as many as possible  snapshot pages between trajectories. 

\vspace{-0.5\baselineskip}
\section{Priority-aware Action Scheduler}
\label{4}

To mitigate the inefficiency of rollout execution discussed in \S\ref{2}, we propose \textit{priority-aware action scheduler}.
We begin by formalizing the optimization problem for end-to-end training.
Subsequently, we propose a length-prediction heuristic to identify high-priority sample groups that contribute to end-to-end training throughput.
We then design priority-aware action scheduler to prioritize high-priority samples across GPU-slot execution, KV Cache residency and request orchestration.

\subsection{Problem Formulation}
\label{sec_problem_formulation_priority}

Let
$
g=\{r_{g,1},\dots,r_{g,K}\}
$
denote a sample group replicated from the same prompt, where $K=1$ for PPO~\cite{schulman2017proximal} and $K>1$ for GRPO~\cite{shao2024deepseekmath}.
A group contributes to training only after all $K$ samples in the group have completed rollout, so its completion time is determined by the slowest sample. 
If one trainer micro-batch requires $B$ ready groups, the optimized training efficiency objective is to minimize the time at which the next $B$ completed groups become available:
\[
\min_{\pi} T_{\mathrm{next}}^\pi(t), \qquad
T_{\mathrm{next}}^\pi(t)=
\inf\{\tau \ge t: |\mathcal{G}_{ready}^\pi(\tau)| \ge B\}.
\]
Here $\pi$ denotes the rollout scheduling policy, and $\mathcal{G}_{ready}^\pi(\tau)$ is the set of groups that have completed rollout by time $\tau$.

This formulation reveals why action-level scheduling adop\allowbreak ted in previous works ~\cite{cao2025skyrl, lu2025iirollflash} alone is insufficient. 
Such a policy improves rollout GPUs utilization, but may delay $T_{\mathrm{next}}^\pi(t)$ since trajectory progress is scattered across many groups, thereby stalling training process.
Moreover, aggressively expanding the set of active trajectories increases KV Cache contention, which may trigger frequent KV Cache eviction, and further introduce reload, or redundant prefill recomputation overhead.
Therefore, the scheduler should prioritize groups that are most valuable for unblocking the next training micro-batch.
We define high-priority groups as the subset of unfinished groups whose completion is expected to most directly reduce $T_{\mathrm{next}}^\pi(t)$.
Once identified, high-priority groups should receive preferential access to GPU slots and KV Cache residency to minimize our optimized goal.

\subsection{Critical Groups Identification}
\label{sec_4_2}

Optimizing training throughput is in fact choosing the best scheduling policy $\pi$ for identifying high-priority groups.
A naive solution is the first-come-first-served (FCFS) policy, which is widely used in scheduling systems.
However, FCFS policy is not the best practice in agentic RL training due to stochastic response dynamics.
Each prompt is replicated into a group of trajectories for advantage estimation.
To encourage exploration and response diversity, a high temperature (e.g., 0.7 for SWE-bench-verified \cite{jimenez2024swebench}) is often set, which increases output-length variance.
Moreover, this variance is further amplified by environment interaction: dynamic and unexpected observations can change the subsequent reasoning path of each trajectory.
As a result, the group arrival order provides limited information about completion order and is insufficient for identifying high-priority groups.

Considering the exponential search space, selecting this critical group set at runtime is fundamentally NP-hard (Detailed proof in Appendix  ~\ref{appendix-np-hard}).
Since complex scheduling logic can itself compromise training throughput, \X instead adopts a lightweight online heuristic.
We propose a length-prediction heuristic policy for identifying critical groups.
The design is based on two observations.

\begin{figure}[t]
    \centering
    \includegraphics[width=0.9\linewidth]{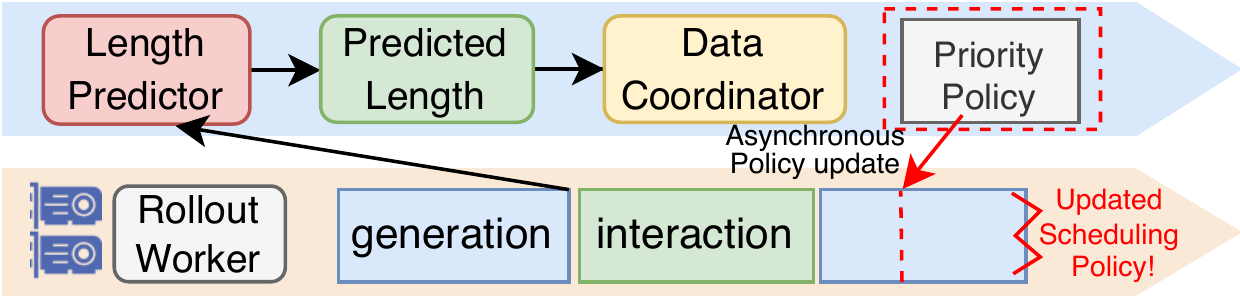}
    \vspace{-0.7\baselineskip}
    \caption[Caption for LOF]{Length-prediction Scheduling Policy.}
    \label{fig:length_predicted_scheduling}
    \vspace{-1.2\baselineskip}
\end{figure}

\textit{\textbf{Observation 1}: per-token latency is nearly constant.}
Action-level scheduling is necessary to keep rollout GPUs saturated.
With the fully asynchronous streaming pipeline, rollout workers continuously receive enough samples for generation with a stable effective batch size.
Thus, the per-token latency is nearly constant and the remaining output length becomes trustworthy for estimating remaining GPU time.

\textit{\textbf{Observation 2}: length prediction accuracy improves as turns accumulate.}
As trajectory history accumulates, it exposes richer signals about the agent's reasoning path, thereby improving the reliability of remaining-length prediction ~\cite{zhang2026heddle}.

Based on these two observations, \X proposes a length-prediction heuristic scheduling policy for identifying high-priority groups.
We collect online trajectory traces and convert each sample into a \texttt{(messages, length)} pair.
The resulting dataset is used to fine-tune a lightweight model, i.e., Qwen3-0.6B.
The model is deployed as a microservice to simplify integration and scale independently from rollout workers.
Meanwhile, to keep prediction off the critical path, \X introduces an \textit{asynchronous priority policy update} mechanism.
As shown in Figure~\ref{fig:length_predicted_scheduling}, each trajectory overlaps length prediction with tool execution.
The prediction result is then asynchronously reported to the data coordinator.
The coordinator updates priorities only when the predicted ordering of sample groups changes.
It then pushes the updated policy to rollout workers without interrupting their generation loops.
Rollout workers continuously perform priority-aware scheduling using the latest available policy.
Therefore, priority decisions do not block the training pipeline or affect the critical path of training progress.

\noindent\textbf{Staleness-boundary promotion.}
Fully asynchronous RL typically enforces a weight-version threshold to bound the maximum number of model updates that any trajectory may span. 
However, length-prediction-based scheduling raises a potential concern that long-running trajectories may be repeatedly demoted, eventually becoming stale enough to be discarded and thereby biasing the training-data stream. 
\X therefore adds a strict safeguard: once a trajectory reaches its last valid weight version, it is immediately promoted to the highest priority until completion, consistent with prior asynchronous RL systems ~\cite{fu2025areal, slime_github} that preferentially serve stale trajectories. 
In practice, fully asynchronous RL often adopts low staleness ratio to preserve training stability; for instance, prior works ~\cite{noukhovitch2025asynchronous, zhong2025streamrl} recommend that accumulated stale samples should not exceed one global batch size.
Such conservative staleness settings limit how long a trajectory can be overtaken by newer requests, substantially restricting priority-induced delay.
We provide a detailed analysis and proof in Appendix ~\ref{appendix-priority-scheduling-reward}, together with experimental evidence in \S\ref{sec:end_to_end_performance}.

\vspace{-0.3\baselineskip}
\subsection{Priority-aware Scheduling}
\label{sec_4_3}

\begin{figure}[t]
    \centering
    \includegraphics[width=0.95\linewidth]{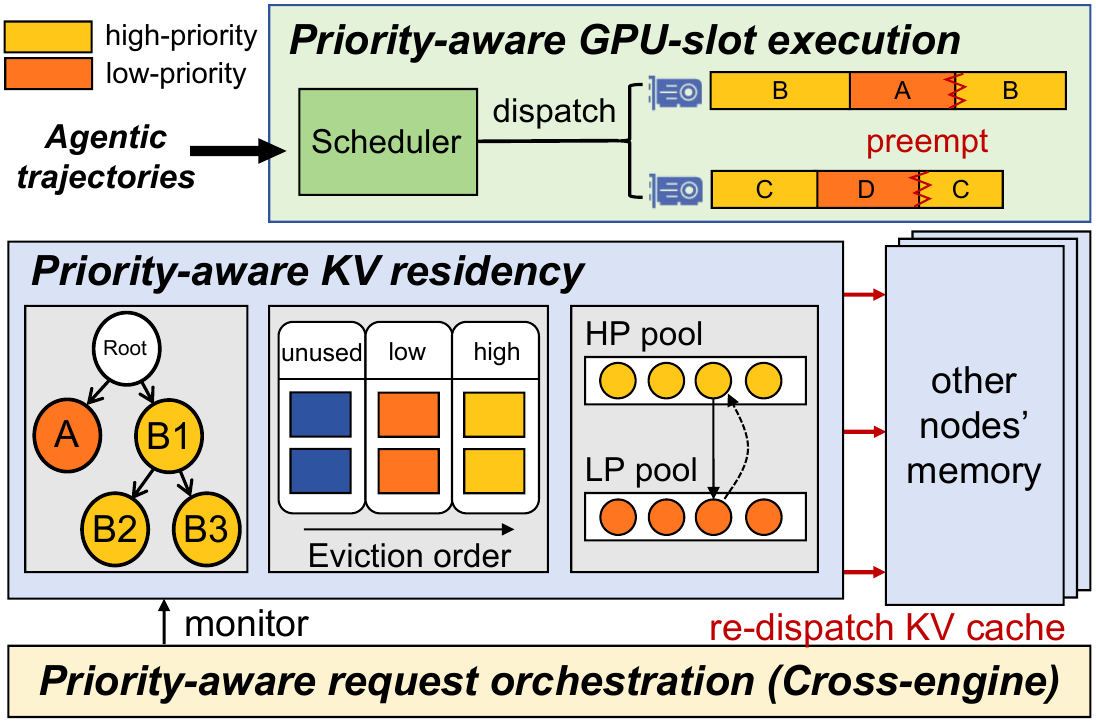}
    \vspace{-0.5\baselineskip}
    \caption[Caption for LOF]{Priority-aware Action Scheduler.}
    \label{fig:priority_aware_action_scheduler}
    \vspace{-1.5\baselineskip}
\end{figure}

Figure~\ref{fig:priority_aware_action_scheduler} presents the overview design of our priority-aware scheduling mechanism.

\noindent\textbf{Priority-aware GPU-slot Execution.}
To preserve the execution order of high-priority samples while maintaining rollout GPU utilization, \X introduces \textit{priority-aware action-level scheduling}.
The scheduler still follows the action-level scheduling paradigm to keep GPU slots saturated.
Simultaneously, instead of scattering GPU slots uniformly across sample groups, it executes low-priority samples opportunistically and high-priority samples preemptively.
Specifically, low-priority samples are admitted only when GPU slots would otherwise be underutilized.
When high-priority samples return from environment interaction and wait for potential GPU slots, the scheduler preempts ongoing low-priority samples.
Their generation histories are preserved as partial rollout states~\cite{zhou2025april}, and the preempted samples are re-enqueued to wait for future GPU slots.

However, enforcing execution priority only at the RL framework is insufficient.
A fixed number of framework-level GPU slots does not perceive contention inside the rollout engine, especially the token-budget contention.
Modern rollout engine ~\cite{kwon2023vllm, zheng2024sglang} typically adopts continuous batching ~\cite{yu2022orca}, where requests are admitted and executed at batch granularity.
Without priority awareness, the rollout engine forms new prefill batches in a FCFS order.
This can cause priority inversion: a high-priority sample may fail to enter the batch due to insufficient token budget, whereas lower-priority samples have already consumed part of the available token budget.
Thus, \X extends preemptive scheduling into the rollout engine.
During prefill batch formation, the rollout engine prioritizes high-priority samples before admitting low-priority ones.
This prevents low-priority samples from occupying token budget  needed by high-priority ones.
Since \X only has two priority classes, batch construction requires only a linear-time partition rather than a full sort, adding negligible scheduling overhead.
For decode batches, when the engine lacks sufficient token space to execute the next forward step, \X retracts low-priority samples first.
The framework-level slot preemption and engine-level priority-aware batching co-design strictly guarantees execution order for high-priority samples.


\noindent\textbf{Priority-aware KV Cache Residency.}
\begin{figure}[t]
    \centering
    \includegraphics[width=1\linewidth]{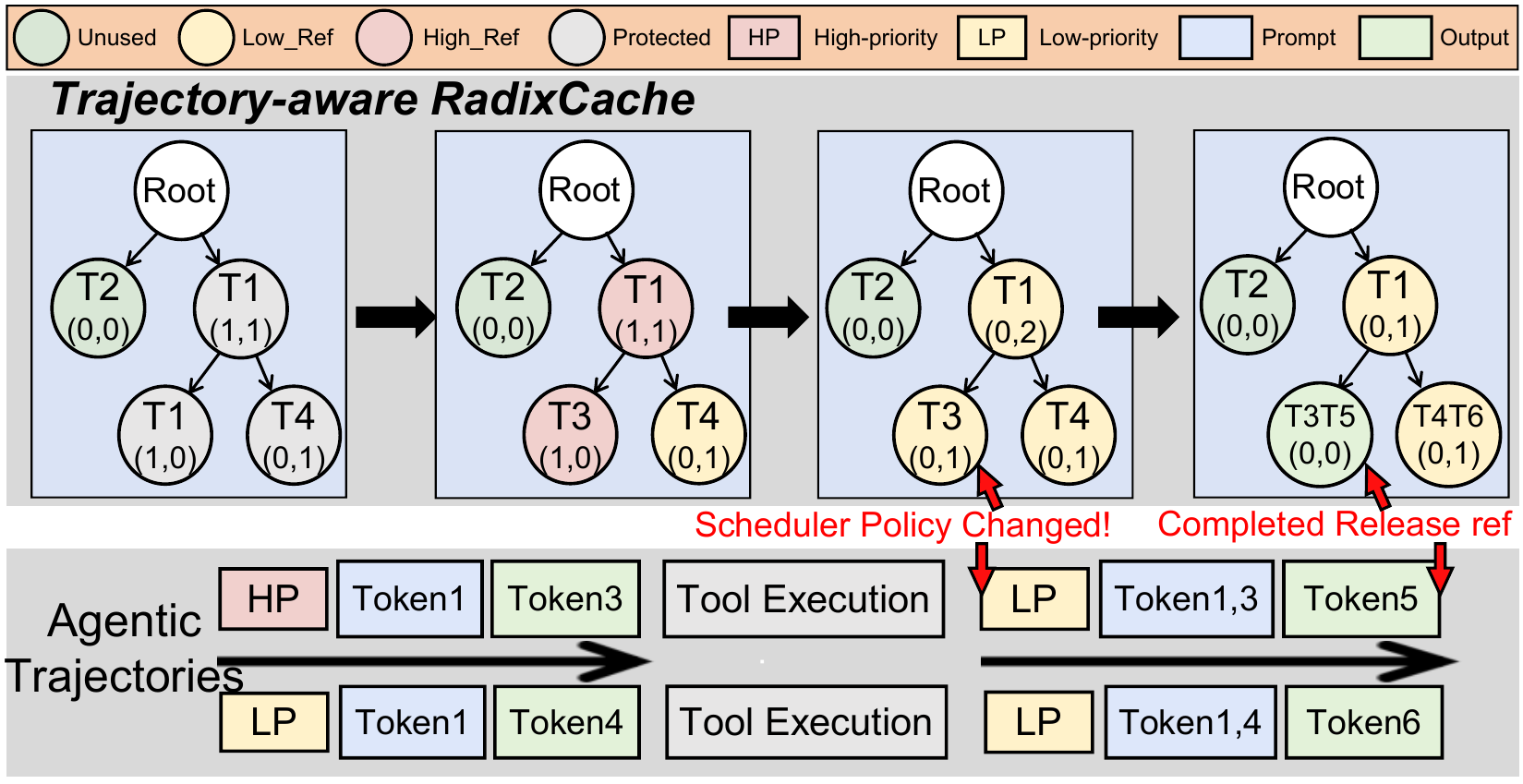}
    \vspace{-1.5\baselineskip}
    \caption[Caption for LOF]{Trajectory-Aware Radix Cache.}
    \label{fig:refAwareHiRadixCache}
    \vspace{-1.2\baselineskip}
\end{figure}
As discussed in \S\ref{2}, the naive KV Cache management mechanism in modern rollout engines ~\cite{kwon2023vllm, zheng2024sglang} introduce frequent KV Cache eviction between multi-turn trajectories.
This contention is further amplified by action-level scheduling due to increasing concurrency of active samples.
Without proactive and priority-aware KV Cache management, high-priority samples may be delayed due to prefix caches reload or recomputation overhead, which compromises training throughput.
To address this issue, \X introduces \textit{trajectory-aware radix cache}, as shown in Figure~\ref{fig:refAwareHiRadixCache}. 
\X retains the prefix-tree architecture used by conventional radix caches, but additionally augments cached token pages with trajectory-level reference counters. 
When each trajectory enters the rollout engine at first time, \X records the priority of the trajectory and adds reference counters on the prefix pages it reuses. 
Unlike request-level protection, this reference is released only when the entire multi-turn trajectory completes.
This allows the cache manager to classify otherwise evictable pages into three tiers: \texttt{unused}, which is not referenced by any inflight trajectory; \texttt{low\_ref}, which is referenced only by low-priority trajectories; and \texttt{high\_ref}, which is referenced by at least one high-priority trajectory.

Building on the trajectory-aware radix cache, \X enforces priority-aware KV Cache residency by using the three-tiered model for both eviction and admission.
When KV pages must be reclaimed for incoming prefill batch or running decode batch, \X evicts pages in tier order: \texttt{unused} first, then \texttt{low\_ref}, and \texttt{high\_ref} last. 
Crucially, the eviction scope is further bounded by the priority of the running requests.
Low-priority requests only reclaim \texttt{unused} and \texttt{low\_ref} tokens, whereas \texttt{high\_ref} pages are considered evictable merely for high-priority requests.
Additionally, \X applies the same tier model at admission time.
For each candidate request, \X computes its residual page budget using only the tiers that the request is permitted to evict.
Crucially, admission and eviction must operate on the same reference snapshot since any skew between the counters used to compute the admission budget and the counters observed during eviction would introduce either memory leak or out-of-memory crashes. 
To enforce this consistency, \X serializes all reference mutations at scheduler-iteration boundaries and applies them atomically before batch formation, preventing any reference update from being interleaved between a batch's admission decision and its KV allocation.
This three-tiered eviction policy prevents low-priority trajectories from evicting KV pages that are still referenced by inflight high-priority trajectories, thereby reducing reload or prefill-recomputation overhead on the critical path of high-priority samples.

However, tiered eviction alone is still insufficient.
Although \X prevents low-priority requests from evicting \texttt{high\_ref} KV pages, severe KV pressure can still cause eviction among high-priority trajectories themselves.
Thus, we propose \textit{resi\allowbreak dency-aware priority demotion}.
Instead of admitting every high-priority request into running batch as long as it can reclaim pages from evictable page pool, \X explicitly constrains the concurrency of high-priority trajectories to retain KV residency.
When the high-priority KV demand exceeds the available KV Cache capacity, \X temporarily selects a group of high-priority inflight trajectories and demotes them to low-priority tier, thereby reserving the KV residency for the remaining high-priority trajectories.
Notably, \X adopts shortest-context-first policy to select demoted trajectories to preserve long trajectories' generation progress.
Once high-priority trajectories complete and release their references, the paused trajectories are promoted back and resumed.
By converting reactive high-priority cache thrashing into controlled priority demotion, this mechanism eliminates repeated eviction among high-priority trajectories and reduces recomputation overhead on the critical path.


\noindent\textbf{Priority-aware Request Orchestration.}
As discussed above, high-priority requests may still be blocked under severe KV contention.
Nevertheless, at the same time, some workers may have sufficient HBM space and execute low-priority samples due to imbalance distribution of high-priority samples.
This creates cross-worker priority inversion.

To address this imbalance, \X proposes \textit{runtime request re-dispatching}.
\X continuously monitors the high-priority KV Cache pressure of each rollout worker.
Once a worker experiences KV contention among high-priority requests, the scheduler redistributes part of its pending requests to workers with the lowest high-priority KV pressure.
To avoid the recomputation overhead of long trajectory, \X migrates the corresponding prefix KV Cache together.
Following prior systems~\cite{zhang2026heddle, zhong2025streamrl}, \X uses Mooncake transfer engine ~\cite{qin2025mooncake} for zero-copy and resilient KV Cache movement.
Notably, to avoid interference with three-tier allocation and admission consistency, \X transfers KV Cache from host memory to host memory rather than directly manipulating on-device HBM cache.
Overall, this design mitigates priority inversion caused by cross-worker load imbalance and improves training throughput.

\vspace{-0.5\baselineskip}
\section{Environment Resource Manager}
\label{5}
\begin{figure}[t]
    \includegraphics[width=0.95\linewidth]{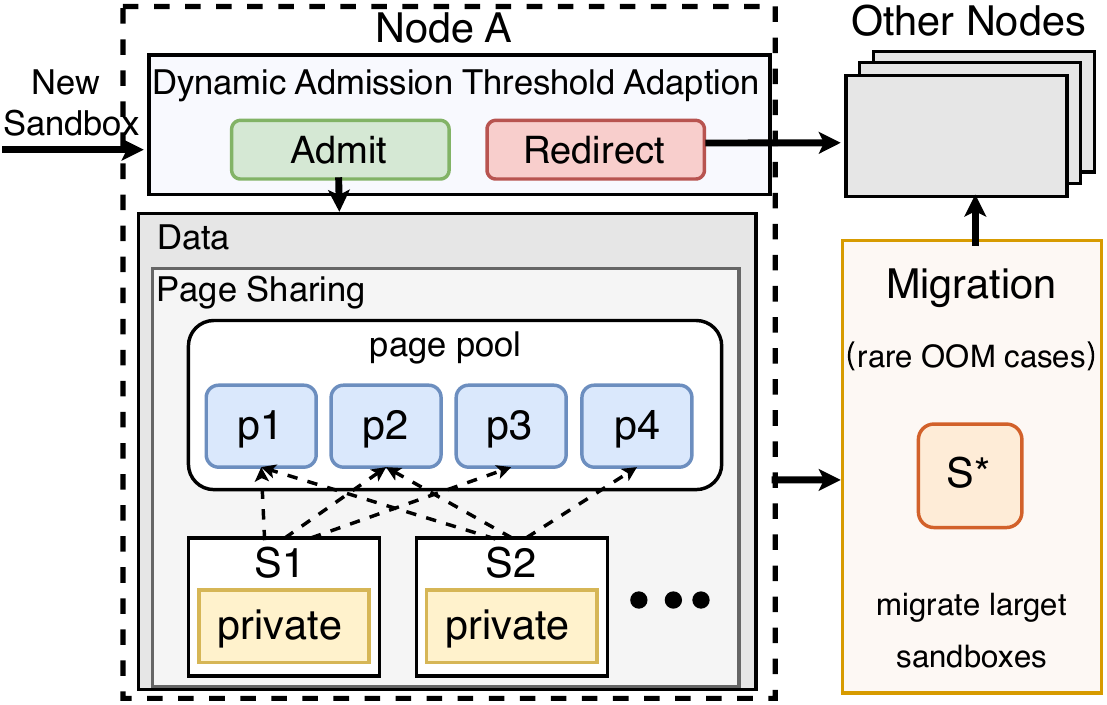}
    \centering
    \vspace{-0.5\baselineskip}
    \caption[Caption for LOF]{Overview of Environment Resource Manager.}
    \label{sandbox}
    \vspace{-1.5\baselineskip}
\end{figure}
In this section, we target the low-density deployment problem of agentic RL sandbox as discussed in \S\ref{2}.
To solve this, we propose page-level sandbox placement design with dynamic admission threshold in \S\ref{51}, and group-level page sharing design in \S\ref{52}.
An overview of the two designs is shown in Figure \ref{sandbox}.

\subsection{Sandbox Placement}
\label{51}

\X first introduces a cluster-level controller to allocate sandboxes from a shared environment pool, enabling high-
density multiplexing across concurrent agentic tasks.
Then, \X performs fine-grained sandbox placement per node to further improve deployment density.
As discussed in \S\ref{2}, coarse-grained admission based on declared sandbox memory leads to severe "memory stranding", where nearly 90\% physical memory is left unused.
After deep dive, we find that E2B ~\cite{e2bsandbox2024} uses Linux \texttt{userfaultfd} to lazily populate snapshot-backed pages, so physical pages are allocated only when they are first touched.
This mechanism explains the gap between actual physical memory usage and virtual upper memory bound, which allows physical node to maintain more sandboxes than permitted.
However, this also creates an admission dilemma. 
If we admit too few sandboxes, it falls back to conservative allocation and wastes physical memory. 
If we admit too many, later page faults, Copy-on-Write (CoW) modification, and runtime allocations may exhaust physical pages and trigger Out-Of-Memory (OOM) issues. 
Therefore, the key problem is to decide how far a node can safely overcommit memory at runtime. 
\X addresses this with a dynamic admission threshold $T$, where details can be seen in Algorithm ~\ref{alg}.
A node admits a new sandbox only when its current physical memory usage is below this threshold.
\X updates it at runtime according to the predicted near-term memory growth. 
Notably, \X uses sandbox migration as a fallback when OOM occurs.

\begin{algorithm}[t]
\caption{Dynamic Admission Threshold Adaptation}
\label{alg}
\begin{algorithmic}[1]
\small
\State \textbf{Input:} $M_{\mathrm{total}}$, $M_{\mathrm{used}}$, $t_{\mathrm{react}}$, $\alpha$, active sandbox set $\mathcal{S}$, candidate sandbox $R$, worker set $\mathcal{W}$
\State \textbf{Output:} admission decision for $R$

\State $G_{\mathrm{active}} \leftarrow \Call{EstimateActiveGrowth}{\mathcal{S}, t_{\mathrm{react}}}$
\State $G_{\mathrm{candidate}} \leftarrow \Call{EstimateColdStartGrowth}{R, t_{\mathrm{react}}}$
\State $H \leftarrow \alpha \cdot \left(G_{\mathrm{active}} + G_{\mathrm{candidate}}\right)$
\State $T \leftarrow 1 - \frac{H}{M_{\mathrm{total}}}$

\If{$M_{\mathrm{used}} < T \cdot M_{\mathrm{total}}$}
    \State \Call{Admit}{$R$}
\Else
    \State \Call{Redirect}{$R$}
\EndIf

\If{\Call{PageAllocationFails}{}}
    \State \Call{InterceptOOM}{}
    \State $\mathcal{S}_{\mathrm{paused}} \leftarrow \mathcal{S}$
    \State \Call{Pause}{$\mathcal{S}_{\mathrm{paused}}$}

    \Repeat
        \State $s, w \leftarrow \Call{SelectSandboxAndWorker}{\mathcal{S}, \mathcal{W}}$
        \State \Call{Migrate}{$s, w$}
        \State $\mathcal{S} \leftarrow \mathcal{S} \setminus \{s\}$
        \State $M_{\mathrm{used}} \leftarrow \Call{RecomputeMemoryUsage}{\mathcal{S}}$
        \State $G_{\mathrm{active}} \leftarrow \Call{EstimateActiveGrowth}{\mathcal{S}, t_{\mathrm{react}}}$
        \State $H \leftarrow \alpha \cdot G_{\mathrm{active}}$
        \State $T \leftarrow 1 - \frac{H}{M_{\mathrm{total}}}$
    \Until{$M_{\mathrm{used}} \leq T \cdot M_{\mathrm{total}}$}

    \State \Call{Resume}{$\mathcal{S}_{\mathrm{paused}}$}
\EndIf

\State \Return
\end{algorithmic}
\end{algorithm}

\noindent\textbf{Dynamic Admission Threshold Adaption.} 
The challenge here is how to estimate future memory growth for determining the admission threshold  $T$.
To answer this, we first deep dive into the memory allocation behavior of Firecracker ~\cite{agache2020firecracker}.
We observe that the physical memory growth mainly comes from four sources: library loading, interpreter runtime caches, allocator operation, and CoW modifications to shared snapshot pages.
A deep investigation shows that sandbox memory growth has a structured lifecycle rather than an arbitrary pattern.
Library loading and interpreter caches dominate the cold-start phase and are largely determined by the sandbox template. 
Allocator operations and CoW modifications continue during running phase, but both shows slower marginal growth than the cold-start phase.
The reason behind is that once allocators have expanded their arenas, subsequent allocations are often served from existing pages.
Similarly, once shared pages have been privatized, later writes to the same pages no longer trigger additional CoW allocation.
As a result, most sandboxes memory growth is typically bursty during cold start and much slower during running phase.
Based on this observation, \X uses two different predict heuristic for cold-start and running sandboxes. 
For cold-start phase sandboxes, \X uses offline profiling of the same template to estimate memory growth, since sandboxes from the same template share the same runtime, dependencies, and initialization path. 
For running sandboxes, \X estimates the recent growth rate from a short time window  and uses this growth rate to estimate their short-term memory demand. 

At admission time, \X accounts for both the recent growth of active sandboxes and the expected cold-start growth of the candidate one.
It admits the candidate only after reserving sufficient headroom for this near-term growth, with a safety factor $\alpha$ to tolerate prediction bias.
When sandbox memory usage is stable, \X raises the threshold and admits more sandboxes; when memory is still growing quickly, \X lowers the threshold and delays further admission.
This mechanism allows \X to increase deployment density while avoiding the unsafe behavior of admitting sandboxes solely based on current memory usage.


\noindent\textbf{Sandbox Migration.}
Although dynamic admission threshold substantially reduces OOM risk, it may still underestimate sudden memory bursts when sandbox footprints grow faster than the predictor. 
To improve robustness, \X uses sandbox migration as a fallback mechanism.

When the controller fails to allocate new physical pages for allocation request, \X intercepts the resulting OOM exception signal before the failure propagates to the task.
It then temporarily pauses all running sandboxes while preserving their in-memory states.
It then selects the sandbox with the largest resident memory footprint and migrates only this sandbox to another resource worker.
After migration, \X recomputes the node's memory usage and admission threshold. 
If the node still exceeds the safe threshold, \X repeats the process. 
This incremental largest-first policy avoids unnecessary migration: \X releases one sandbox's memory at a time and stops as soon as the node returns to a safe state.
To reduce migration overhead and limit its impact on the critical path, \X uses the Mooncake transfer engine ~\cite{qin2025mooncake} to transfer sandbox memory state directly between resource workers. 
This direct memory-state transfer avoids slow checkpoint-and-reload paths and allows \X to quickly relieve memory pressure while preserving sandbox execution state.
Upon migration, \X resumes all paused running sandboxes.

\vspace{-1\baselineskip}
\subsection{Page Sharing}
\label{52}

\begin{figure}[t]
    \includegraphics[width=0.95\linewidth]{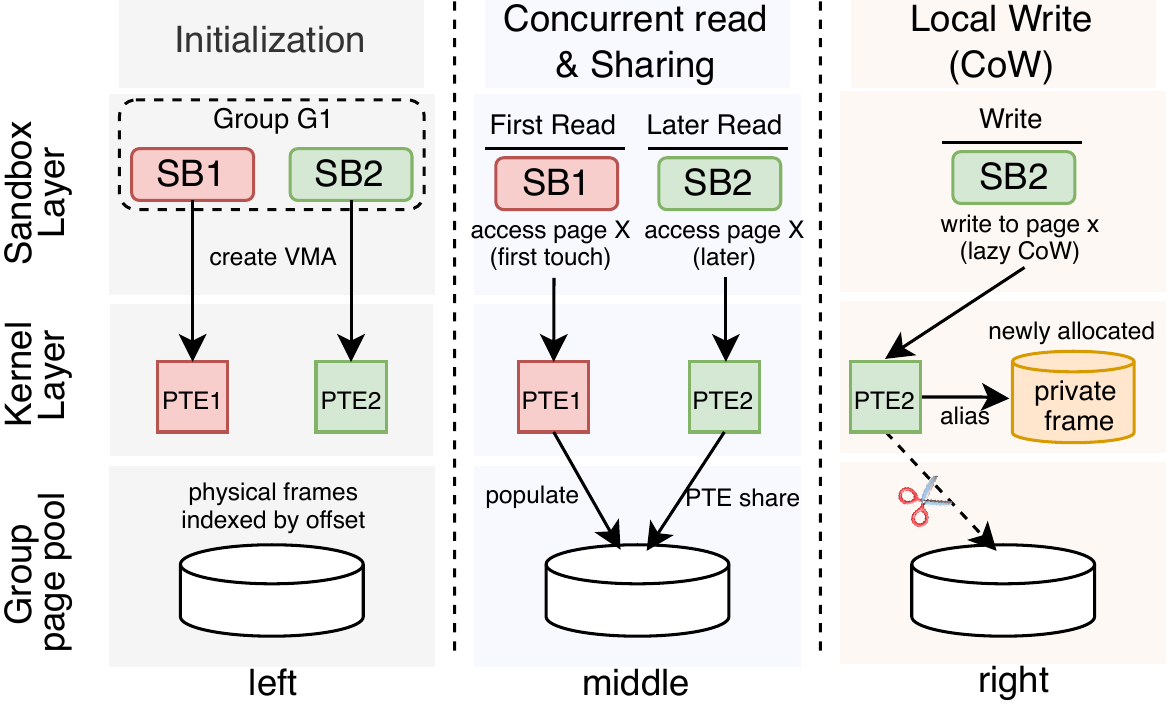}
    \centering
    \vspace{-1\baselineskip}
    \caption[Caption for LOF]{Mechanism of Page Sharing.}
    \label{sandbox2}
    \vspace{-1\baselineskip}
\end{figure}

RL algorithms such as PPO~\cite{schulman2017proximal} and GRPO~\cite{shao2024deepseekmath} expand each prompt into a group of trajectories that, by construction, are launched from the same environment template. Within a group, sandboxes execute nearly identical setup code—loading the same dependencies, importing the same packages, materializing the same scratch files—so a large fraction of their pages remain byte-identical until divergent tool calls begin to write new state. \X exploits this invariant to increase per-node sandbox density without weakening isolation.
As shown in Figure \ref{sandbox2}, \X realizes intra-group sharing by decoupling three issues that the native fork/mmap path conflates, i.e., virtual-address ownership, physical-page residency, and write permission. Each sandbox keeps its own VMA (Virtual Memory Area) and page table, but read-only PTEs (Page Table Entry) from different sandboxes in a group may alias the same physical frame (shared residency), with write permission masked until a write actually occurs, i.e., lazy CoW.
Two requirements must be satisfied. (i) Eager binding: sharing should be installed at sandbox creation, before any page fault, so that no sandbox ever allocates a redundant frame for a snapshot page (Figure \ref{sandbox2}-left). (ii) Strict isolation: a write by one sandbox can't be visible to its peers and compromised sandboxes can't corrupt the shared backing store. 
\X realizes this across two layers: a user-space group page pool and a kernel-side write-protected mapping.

\noindent\textbf{Group Page Pool At User Space.} \X maintains a per-group page pool indexed by guest-memory offset, using the environment template as the sharing key. 
Pages are populated lazily: when any sandbox in the group first accesses a page, \X loads that page into the shared pool (Figure \ref{sandbox2}-middle). 
For each memory offset, \X allows only one sandbox to perform this first load at a time, so concurrent sandboxes always share the same populated page. 
Once populated, a pool entry is immutable: subsequent accesses from any group member find the frame already resident, with no per-sandbox allocation. 
Because the pool is keyed by template rather than by content hash, \X never scans memory to discover sharing; every read-only snapshot page is shared by construction.

\noindent\textbf{Per-Sandbox Mapping And CoW At Kernel Level.} Each sandbox maps its guest memory as a private, write-protected view over the page pool. Reads behave as ordinary guest-memory accesses, with the sandbox's PTE silently aliasing a frame in the pool. A write is handled by allocating a private writable frame, copying the page contents, and rebinding only that sandbox's PTE to the new frame.
The pool entry and every other sandbox's mapping are left untouched (Figure \ref{sandbox2}-right). 
Because aliasing is performed at PTE granularity rather than by sharing whole memory regions, divergence is local---isolation is preserved at the same granularity that the sandbox boundary already enforces, and a compromised sandbox can corrupt only its own private frames.
\vspace{-0.5\baselineskip}
\section{Evaluation}
\label{sec:evaluation}
\begin{figure*}[t!]
    \ifarxiv
        \includegraphics[width=0.95\linewidth]{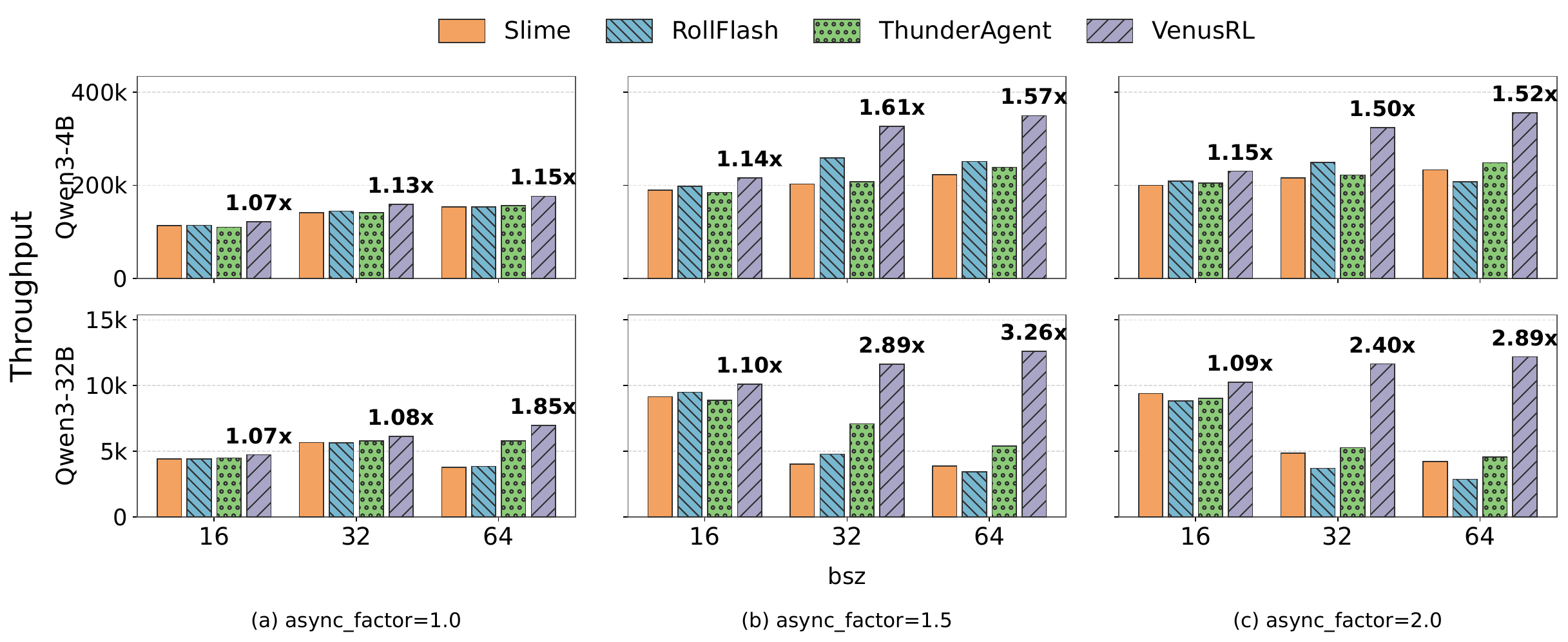}
    \else
        \includegraphics[width=0.95\linewidth]{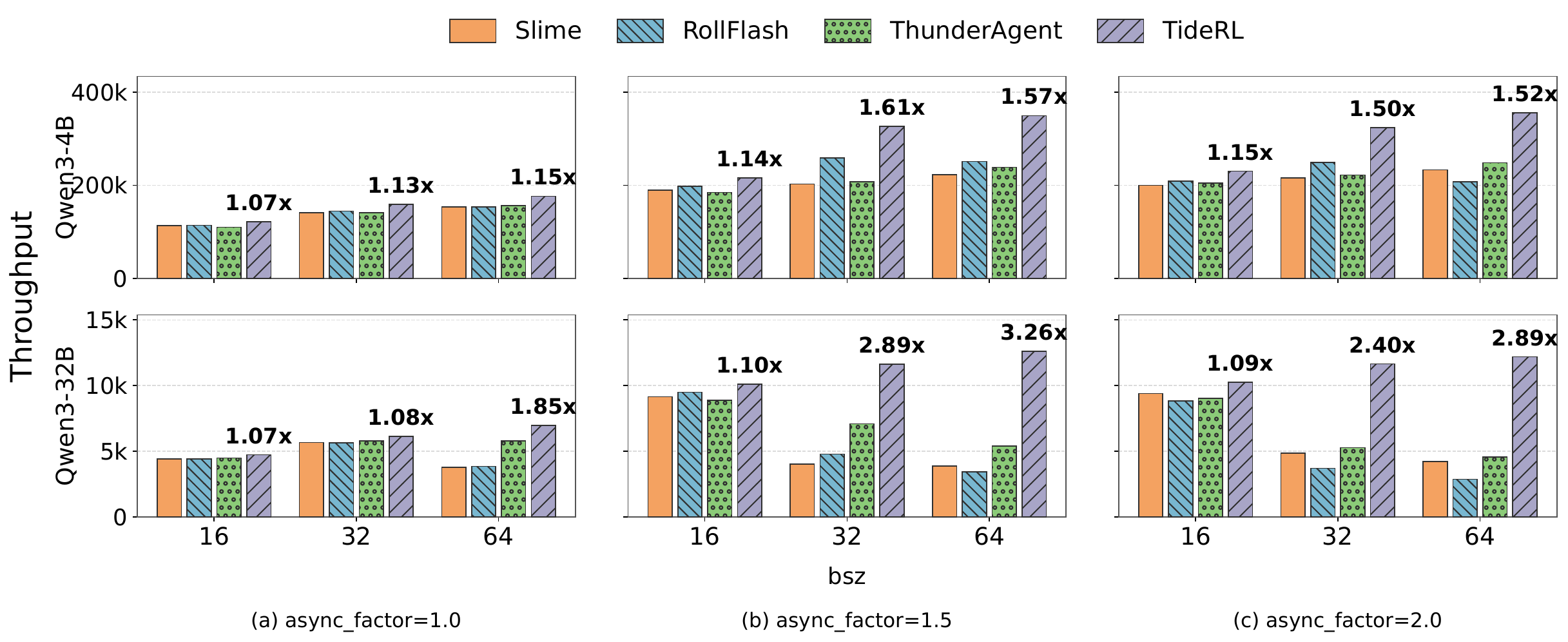}
    \fi
    \centering
    \vspace{-0.5\baselineskip}
    \caption[Caption for LOF]{Training performance of agentic RL systems under different batchsize and staleness.}
    \label{fig:training_performance}
    \vspace{-0.5\baselineskip}
\end{figure*}
We implemented \X with 24k lines of Python, Rust and Golang atop our internal team's RL training framework, SGLang ~\cite{zheng2024sglang}, E2B ~\cite{e2bsandbox2024} and Firecracker ~\cite{agache2020firecracker}.
Our evaluation aims to answer the following questions:
(1) How does \X compare to state-of-the-art agentic RL systems in end-to-end training performance (\S\ref{sec:end_to_end_performance})? (2) How do the three core techniques in the priority-aware action scheduler contribute to performance (\S\ref{sec:abalation_study_action_scheduler})? (3) How effective is the environment resource manager in \X (\S\ref{sec:Effectiveness_environment})?

\vspace{-0.5\baselineskip}
\subsection{Experimental Setup}
\noindent\textbf{Testbed.}
We conduct our experiments on a cluster of four servers, totaling 32 GPUs.
Each node is equipped with eight NVIDIA Hopper GPUs, 180 Intel Xeon CPU cores, and 1.8 TB of host memory. 
Intra-node GPU communication utilizes 900 GB/s NVLink, while inter-node networking is handled by 400 Gb/s NVIDIA Mellanox InfiniBand.
Specifically, we allocate 16 GPUs to generation and another 16 GPUs to training.

\noindent\textbf{Models and Workloads.}
We evaluate \X using instruc\allowbreak tion-tuned Qwen3-4B and Qwen3-32B models ~\cite{bai2023qwen}, which provide strong tool-use and reasoning capabilities.
We conduct experiments on representative SWE-agent tasks ~\cite{yang2024swe} with the OpenSWE dataset ~\cite{fu2026davincienvopensweenvironment}, which contains 45,320 executable Docker environments spanning over 12.8K repositories, allowing agents to execute code, run test suites, and check formatting.
For all agentic rollouts, we set the maximum output length to 128K tokens and limit each trajectory to at most 300 turns.
We generate eight samples per prompt using the GRPO RL algorithm ~\cite{shao2024deepseekmath}.
We set the LLM generation temperature to 0.7 ~\cite{qwen2} and configure the framework with a weight-version threshold of 2, discarding samples whose generation spans three or more weight versions.

\noindent\textbf{Baselines.}
We compare \X against three state-of-the-art agentic RL systems:
\noindent(1) \textit{Slime} ~\cite{slime_github} is one of the most widely used open-source frameworks for agentic RL training. 
It adopts trajectory-level scheduling and may suffer from turn-level agentic bubbles.
\noindent(2) \textit{RollFlash} ~\cite{lu2025iirollflash} disaggregates environment controller from rollout controller, adopting action-level scheduling policy. 
It allows one sample's tool execution to overlap with the another sample's generation, thereby improving GPU utilization.
\noindent(3) \textit{ThunderAgent} ~\cite{kang2026thunderagent} abstracts agentic workflows as programs and adopts a program-aware scheduler that mitigates KV Cache thrashing by proactively evicting shorter contexts under memory pressure.

Notably, given that the SWE-agent workload produces extremely long responses under our setting, we enable the hierarchical cache (HiCache) in SGLang ~\cite{hicache} to support offloading KV Cache to L2 host memory.
To ensure fair comparison, we bump the sglang version to 0.5.17 and keep all the framework or algorithm arguments consistent.

\subsection{End-to-End Performance}
\label{sec:end_to_end_performance}

\noindent\textbf{Training performance.}
We first benchmark \X against three state-of-the-art agentic RL systems on training performance across different settings.
The \texttt{async\_factor} set here controls the maximum staleness samples allowed in each rollout step.
In particular, setting \texttt{async\_factor} to 1.0 effectively enforces strict on-policy RL training.

\begin{figure}[t]
    \ifarxiv
        \includegraphics[width=1\linewidth]{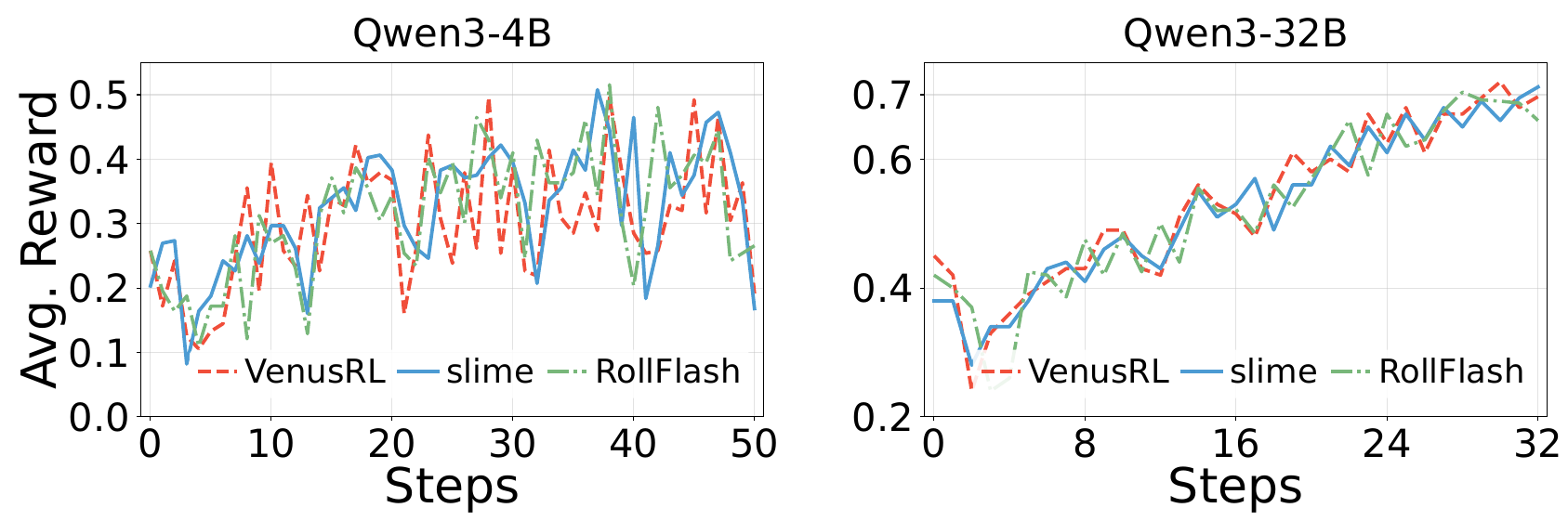}
    \else
        \includegraphics[width=1\linewidth]{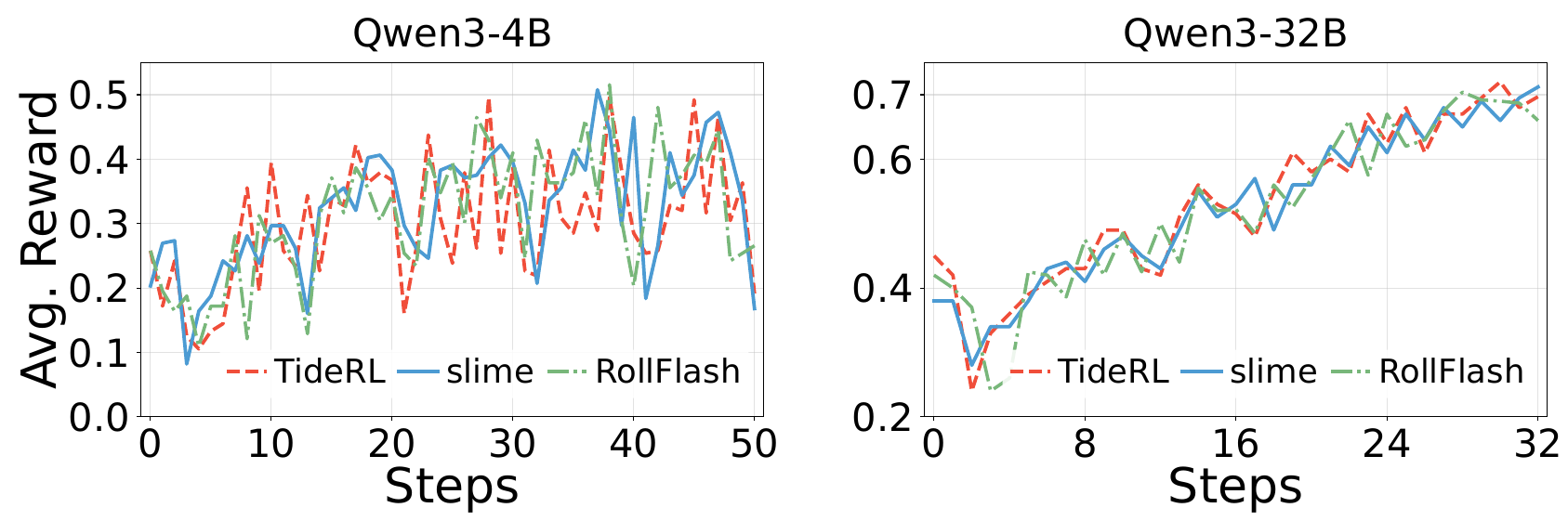}
    \fi
    \centering
    \vspace{-1.5\baselineskip}
    \caption[Caption for LOF]{Reward score for training.}
    \label{fig:reward_curves}
    \vspace{-1\baselineskip}
\end{figure}

\begin{figure}[t]
    \ifarxiv
        \includegraphics[width=1\linewidth]{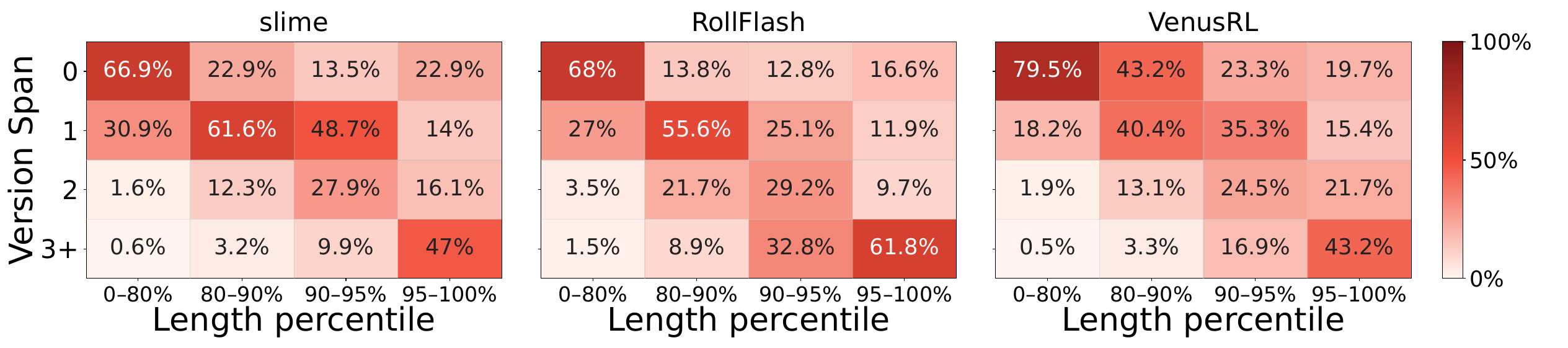}
    \else
        \includegraphics[width=1\linewidth]{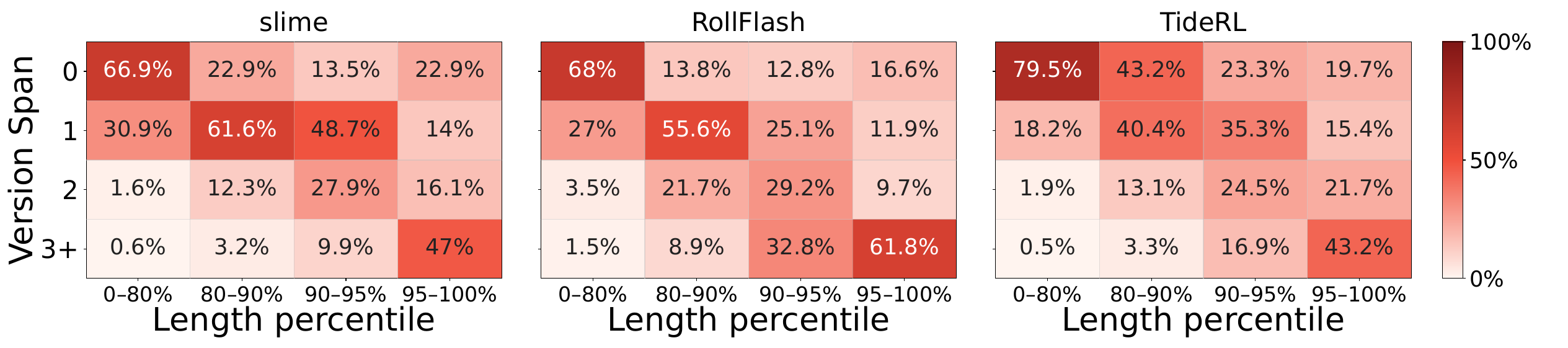}
    \fi
    \centering
    \vspace{-1.5\baselineskip}
    \caption[Caption for LOF]{Weight version heatmap comparison.}
    \label{fig:weight_version_heatmap}
    \vspace{-1\baselineskip}
\end{figure}

\noindent \underline{1) End-to-End Training Speed.}
Figure ~\ref{fig:training_performance} presents the end-to-end training throughput across various batch sizes and \texttt{async\_factor} configurations.
Across both models, \X consistently outperforms all three baselines across all settings, achieving 1.07--3.26$\times$ speedup over Slime, 1.06-4.24$\times$ speedup over RollFlash, and 1.05--2.67$\times$ speedup over ThunderAgent.
Even under strict on-policy training, i.e., \texttt{async\_fa\allowbreak ctor}=1.0, \X improves throughput in every setting: by 1.07--1.15$\times$ for Qwen3-4B and 1.05--1.85$\times$ for Qwen3-32B.
This gain mainly comes from the priority-aware KV Cache residency, which prioritizes evicting unused KV pages before evicting prefix pages referenced by in-flight trajectories.
In addition, runtime request orchestration balances the sample distribution across rollout workers, further explaining the gain.
When \texttt{async\_factor} increases to 1.5 or 2.0, the speedup at batch size 16 remains modest at 1.06--1.17$\times$, consistent with limited KV cache contention and fewer opportunities for priority-aware scheduling.
At batch sizes 32 and 64, \X achieves 1.26--1.71$\times$ speedup for Qwen3-4B and 1.64--4.24$\times$ for Qwen3-32B.
The experimental results demonstrate the strong scalability of \X for agentic RL training.

\noindent \underline{2) Convergence and Stability.}
Figure ~\ref{fig:reward_curves} presents the reward score of Slime, Rollflash and \X under 1.5 \texttt{async\_factor} and 32 batch size setting.
All three systems converge to nearly identical rewards within comparable steps.
We observe similar convergence behavior across all evaluated settings; due to space limitations, we only report this representative setting.

Figure ~\ref{fig:weight_version_heatmap} shows weight-version heatmaps across trajectory length percentiles, accumulated over 100 training steps with Qwen3-4B. 
\X incurs no higher long-tail discard rate than Slime or Rollflash, justifying our design choice. 
Notably, we observe higher long-tail discard rate in Rollflash, which stems from action-level scheduling.
Without special handling, stale trajectories repeatedly compete with newer
\newpage
\noindent samples and their progresses are continuously blocked.

\begin{figure}[t!]
    \centering
    \begin{minipage}[b]{0.48\linewidth}
        \centering
        \ifarxiv
            \includegraphics[width=\linewidth]{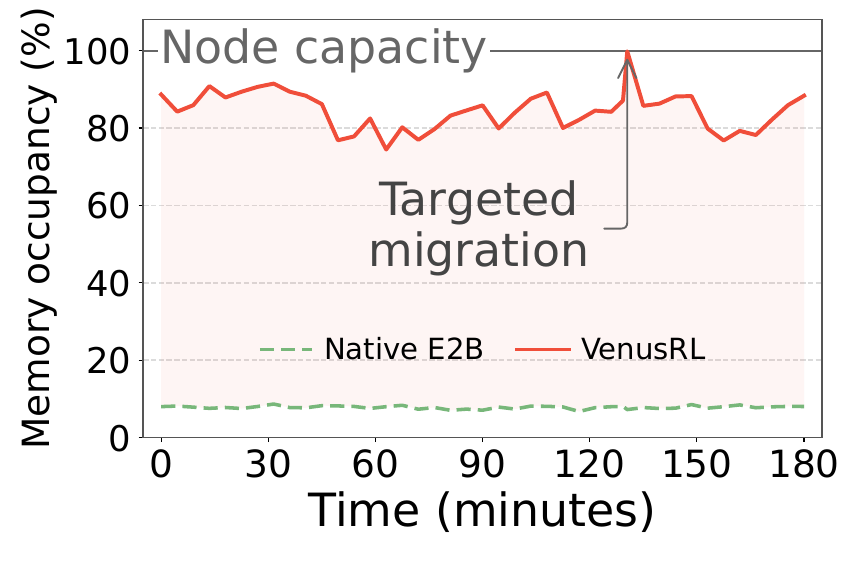}
        \else
            \includegraphics[width=\linewidth]{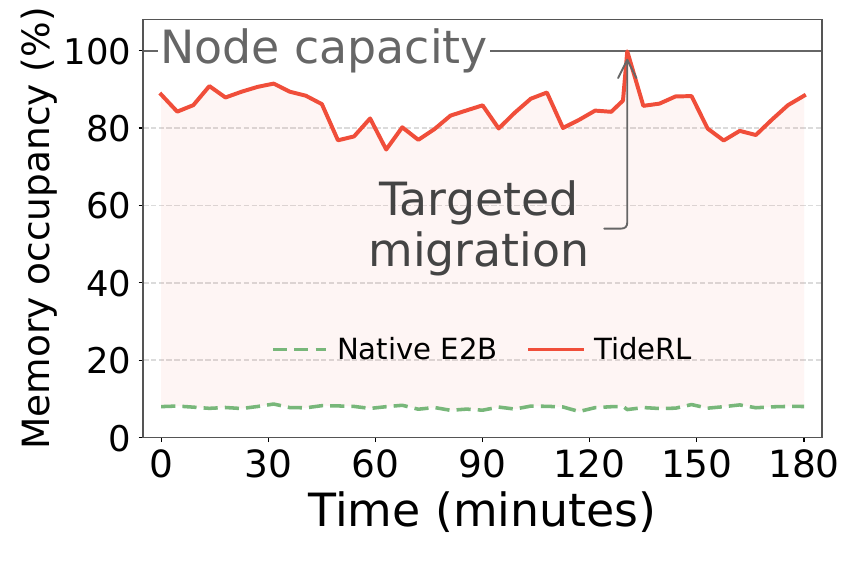}
        \fi
        \vspace{-2\baselineskip}
        \caption{Online sandbox Memory occupancy.}
        \label{fig:occupancy}
    \end{minipage}
    \hfill
    \begin{minipage}[b]{0.48\linewidth}
        \centering
        \ifarxiv
            \includegraphics[width=\linewidth]{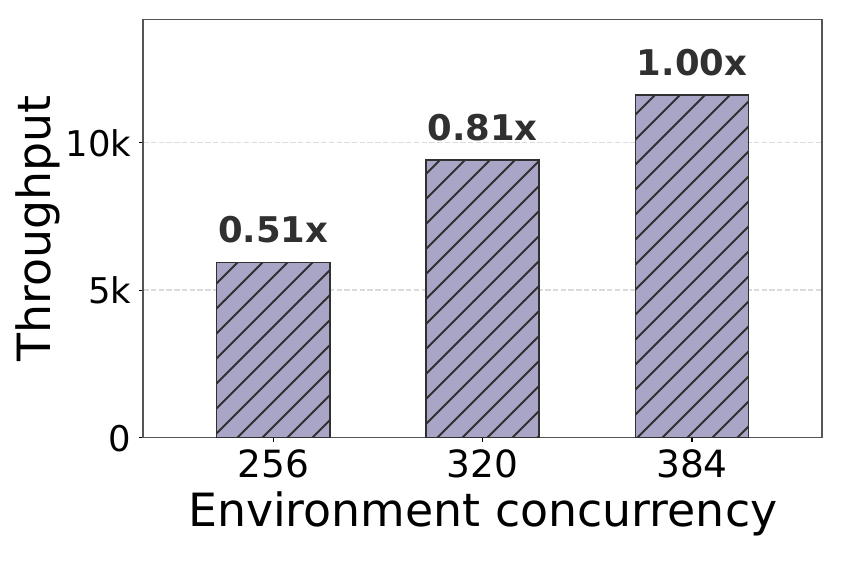}
        \else
            \includegraphics[width=\linewidth]{Figure/plot/main/environment_concurrency.pdf}
        \fi
        \vspace{-2\baselineskip}
        \caption{Throughput under limited resources.}
        \label{fig:resource_constrained_throughput}
    \end{minipage}
    \vspace{-1\baselineskip}
\end{figure}

\noindent\textbf{Environment management.} 
As discussed in \S\ref{2}, memory capacity is the primary bottleneck that caps sandbox deployment density.
Figure~\ref{fig:occupancy} thus profiles node memory occupancy over a three-hour online run case on a 400GB node. 
Native E2B admits sandboxes by configured peak memory. 
Occupancy saturates at about $8\%$ and stays flat for the rest of the run, leaving over $90\%$ of node memory unusable.
In contrast, \X admits sandbox based on  actual memory usage and keeps occupancy in a high band between roughly 80\% and 95\%.
At minute~132, a burst-growth of sandbox footprint drives the node occupancy to the capacity line within a single control interval.
Rather than surrendering the node to the kernel OOM killer, \X triggers targeted sandbox migration and returns occupancy in under 3\,seconds (\S\ref{sec:Effectiveness_environment}).
In production, such issue is near zero with a well-tuned admission-threshold (specified in Appendix  ~\ref{appendix-OOM}).
We deliberately select a window that contains one such event to show how \X behaves under residual OOM pressure.


Figure ~\ref{fig:resource_constrained_throughput} evaluates how environment concurrency backpressures rollout efficiency when sandbox resources are constrained, with 32 batch size and 1.5 \texttt{async\_factor} setting.
As is depicted, throughput drops to 0.81$\times$ at 320 concurrent sandboxes and to 0.51$\times$ at 256, respectively.
The data indicates that without high-density sandbox deployment, the limited sandbox resources may directly degrade training performance and hide the efficiency of priority-aware scheduler.


\subsection{Priority-Aware Action Scheduler}
\label{sec:abalation_study_action_scheduler}

\begin{table}[t]
    \centering
    \caption{End-to-end training speedup breakdown.}
    \vspace{-1\baselineskip}
    \label{tab:training_speedup_breakdown}
    \footnotesize
    \setlength{\tabcolsep}{5pt}
    \renewcommand{\arraystretch}{1.15}
    \resizebox{\columnwidth}{!}{
    \begin{tabular}{lccc}
        \toprule
        \textbf{Method}
        & \textbf{\makecell{bsz=16}}
        & \textbf{\makecell{bsz=32}}
        & \textbf{\makecell{bsz=64}} \\
        \midrule
        Baseline               & 1.00 & 1.00 & 1.00 \\
        + priority-aware schedule       & 1.01$\times$ & 1.81$\times$ & 1.42$\times$ \\
        + kvcache residency    & 1.06$\times$ & 2.72$\times$ & 2.78$\times$ \\
        + request migration    & 1.10$\times$ & 2.89$\times$ & 3.26$\times$ \\
        \bottomrule
    \end{tabular}
    }
    \vspace{-2\baselineskip}
\end{table}

\noindent\textbf{Performance breakdown.}
    Table ~\ref{tab:training_speedup_breakdown} presents a detailed breakdown of the cumulative speedup from our proposed techniques in priority-aware action scheduler across different batch sizes with Qwen3-32B and fixed \texttt{async\_factor} of 1.5.
\begin{itemize}[leftmargin=*]
\item Priority-aware scheduling improves training throughput by prioritizing sample groups that are most likely to unblock the next trainer batch. 
Its benefit becomes more pronounced as the batch size increases, achieving 1.81$\times$ and 1.42$\times$ speedup at batch sizes of 32 and 64, respectively. 
Notably, the improvement decrease from bsz 32 to 64 indicates that without priority-aware KV Cache residency, the scheduling policy may not fully unleash its potential.

\item KV Cache residency further improves performance by reducing unnecessary KV Cache eviction and recomputation for high-priority samples. 
This component provides substantial gains across all batch sizes, providing a 1.36$\times$ performance uplift (from 1.42$\times$ to 2.78$\times$) at batch size 64. 
It also retains its gains for small batch sizes, contributing a 0.91$\times$ improvement (from 1.81$\times$ to 2.72$\times$) at batch size 32.

\item Request migration provides additional gains by rebalancing high-priority rollout requests across workers and reducing KV contention. 
In particular, it delivers up to 0.48$\times$ (from 2.78$\times$ to 3.26$\times$) speedup for batch size of 64 setting.

\end{itemize}

\noindent\textbf{Ablation study.}
We further examine whether the key design choices inside the priority-aware action scheduler are necessary and optimized with Qwen3-32B.

\noindent\underline{1) Length-prediction scheduling policy.}
\begin{figure}[t]
    \ifarxiv
        \includegraphics[width=0.95\linewidth]{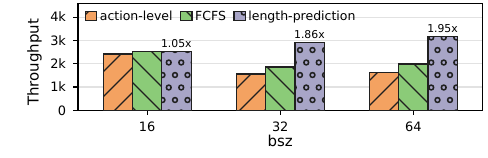}
    \else
        \includegraphics[width=0.95\linewidth]{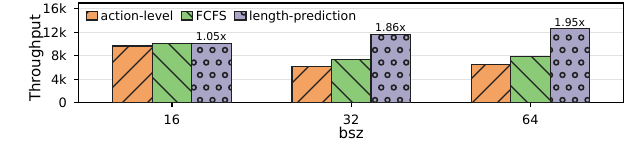}
    \fi
    \centering
    \vspace{-1\baselineskip}
    \caption[Caption for LOF]{Scheduling policy comparison.}
    \label{fig:ablation_rollout_scheduling}
    \vspace{-1\baselineskip}
\end{figure}
We first evaluate the effectiveness of our length-prediction scheduling policy.
In this experiment, we keep the other system components enabled, fix \texttt{async\_factor} to 1.5, and compare three different scheduling policies.
As shown in Figure~\ref{fig:ablation_rollout_scheduling}, the length-prediction policy consistently outperforms the other policies across all batch sizes.
It achieves up to 95\% higher end-to-end training throughput than action-level policy, and up to 58\% than FCFS-based priority policy.

\begin{figure}[t!]
    \centering
    \begin{subfigure}[b]{0.48\linewidth}
        \centering
        \includegraphics[width=\linewidth]{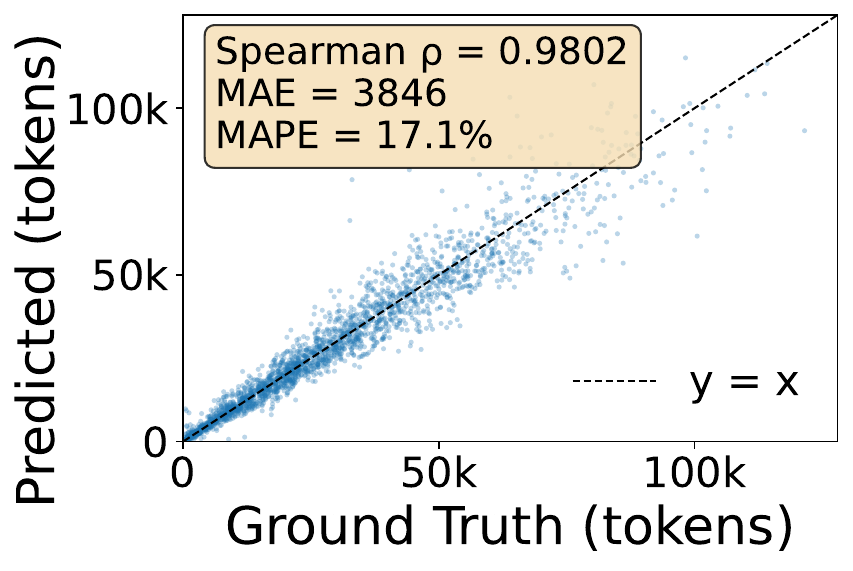}
        \vspace{-1\baselineskip}
        \caption{Predictor accuracy}
        \label{fig:predictor_accuracy}
    \end{subfigure}
    \hfill
    \begin{subfigure}[b]{0.48\linewidth}
        \centering
        \includegraphics[width=\linewidth]{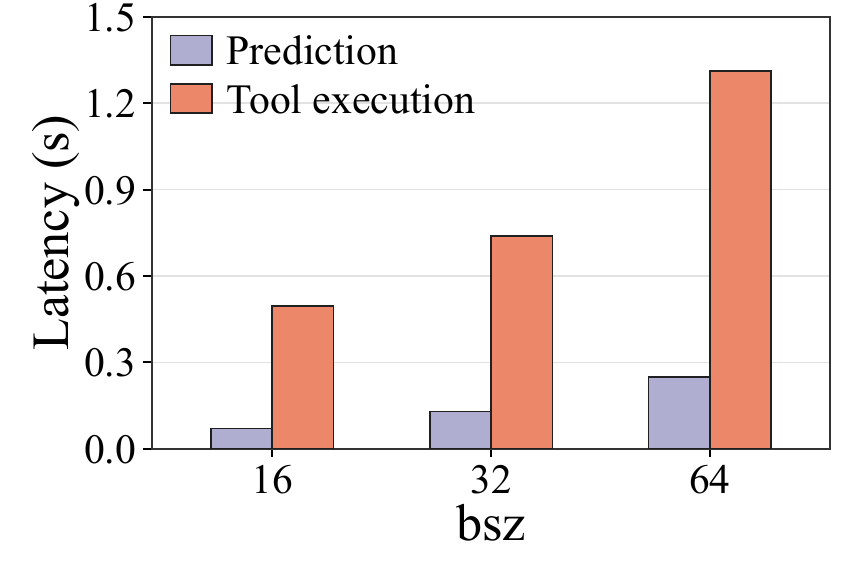}
        \vspace{-1\baselineskip}
        \caption{Predictor overhead}
        \label{fig:predictor_latency}
    \end{subfigure}
    \vspace{-1\baselineskip}
    \caption{Predictor performance.}
    \label{fig:predictor_performance}
    \vspace{-1\baselineskip}
\end{figure}
We then profile the runtime behavior of the length predictor.
As shown in Figure ~\ref{fig:predictor_accuracy}, the predictor achieves a Spearman correlation ~\cite{spearman1930c} of 0.98, confirming that it reliably distinguishes long requests from short ones—sufficient for priority-based scheduling.
Though absolute generation lengths may shift as the rollout policy evolves, the predictor conditions on the problem specification and accumulated trajectory history to capture task difficulty and execution progress, providing informative cues for relative ordering.
Additionally, the predictor can also be updated at training checkpoints using recent trajectories to mitigate policy drift, with fine-tuning typically converging within only minutes (More details in Appendix ~\ref{appendix-predictor-robustness}).
On the other hand, Figure ~\ref{fig:predictor_latency} reports the average predictor latency and corresponding tool interaction time.
The predictor latency is much smaller than the interaction latency, thus the overhead can be effectively overlapped.

\noindent\underline{2) Priority-aware KV Cache residency.}
\begin{figure}[t!]
    \centering
    \vspace{-0.5\baselineskip}
    \begin{subfigure}[b]{0.48\linewidth}
        \centering
        \ifarxiv
            \includegraphics[width=\linewidth]{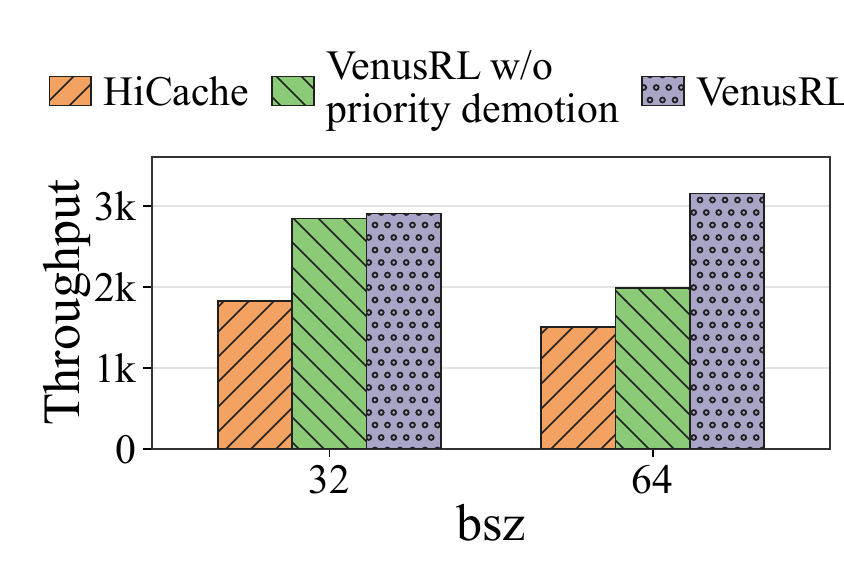}
        \else
            \includegraphics[width=\linewidth]{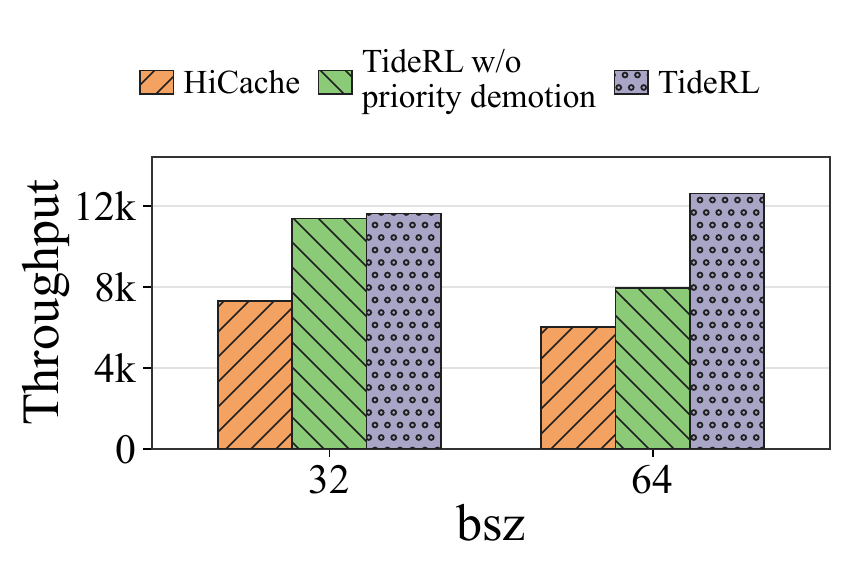}
        \fi
        \vspace{-1.5\baselineskip}
        \caption{Training performance}
        \label{fig:ablation_kvcache_performance}
    \end{subfigure}
    \hfill
    \begin{subfigure}[b]{0.48\linewidth}
        \centering
        \ifarxiv
            \includegraphics[width=\linewidth]{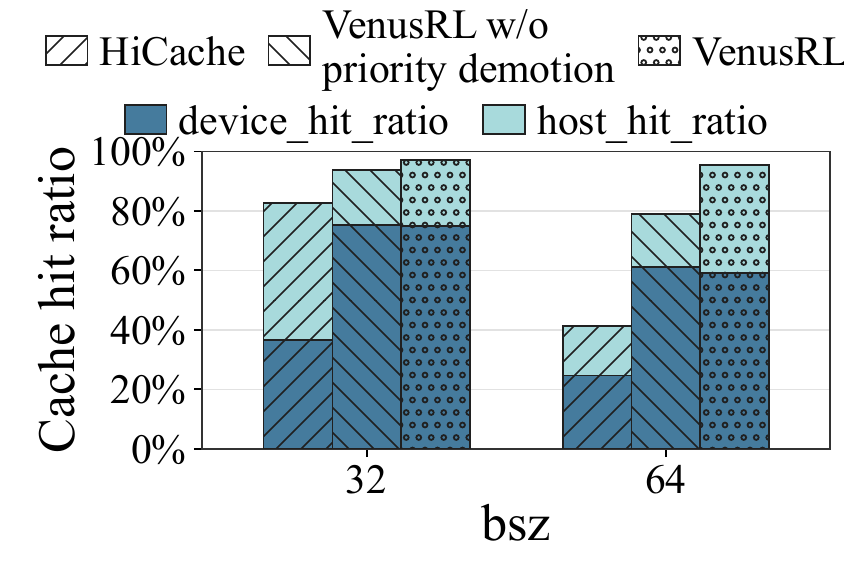}
        \else
            \includegraphics[width=\linewidth]{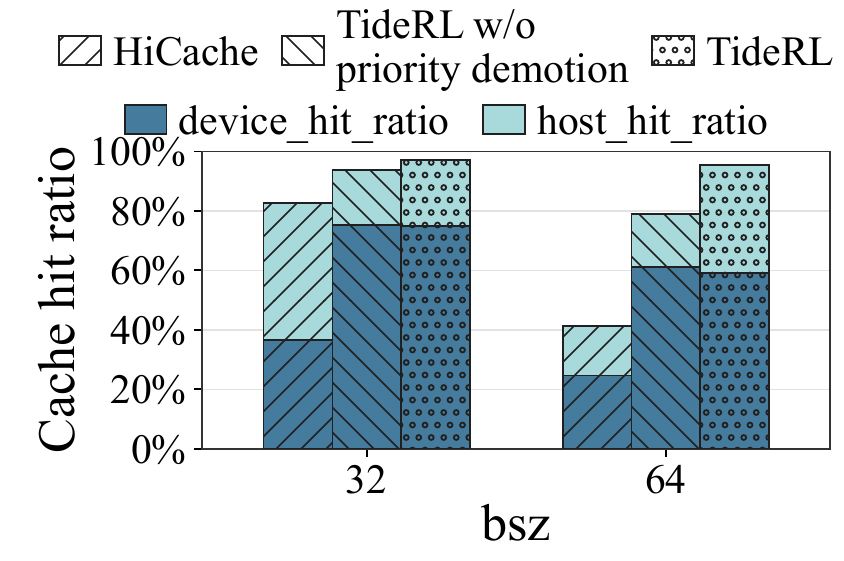}
        \fi
        \vspace{-1.5\baselineskip}
        \caption{Cache hit ratio}
        \label{fig:ablation_cache_hit_ratio}
    \end{subfigure}
    \vspace{-1\baselineskip}
    \caption{Effectiveness of KV Cache residency.}
    \label{fig:ablation_kvcache_residency}
    \vspace{-1.1\baselineskip}
\end{figure}
We next evaluate how different components in \X's priority-aware KV Cache residency affect training performance.
Figure ~\ref{fig:ablation_kvcache_performance} and Figure ~\ref{fig:ablation_cache_hit_ratio} compare three cache backends across different batch sizes: HiCache, \X’s trajectory-aware radix cache with LRU eviction for high-priority samples under memory pressure, and full \X with residency-aware priority demotion.
Compared with HiCache, the trajectory-aware radix cache improves end-to-end training throughput by preserving KV Cache state of inflight trajectories and thus increasing the cache hit rate.
It maintains a 93\% cache hit rate under a batch size of 32, compared with 82\% for HiCache, and achieves a 56\% speedup.
However, the cache hit rate drops to only 79\% under the batch size of 64 due to severe KV Cache pressure, achieving sub-optimal performance.
\X mitigates this degradation with residency-aware priority demotion, maintaining an over 95\% cache hit rate under the same batch size.
As a result, \X achieves up to 2.09$\times$ speedup over HiCache and 1.58$\times$ speedup over \X without priority demotion.
The experimental result justifies \X's design choices of priority-aware KV Cache residency.

\subsection{Environment Resource Manager}
\label{sec:Effectiveness_environment}

\begin{figure}[t!]
    \centering
    \begin{minipage}[b]{0.48\linewidth}
        \centering
         \ifarxiv
            \includegraphics[width=\linewidth]{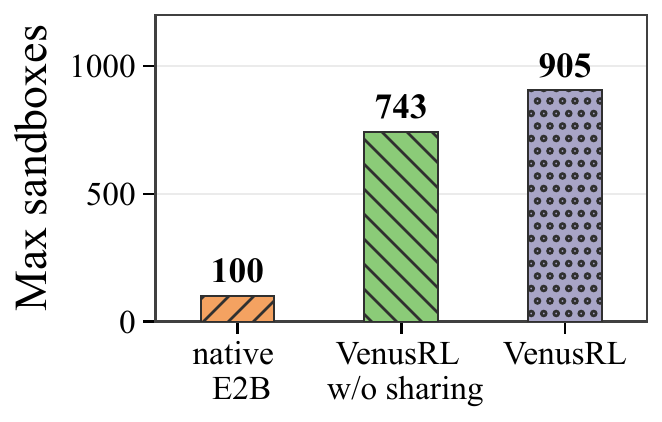}
        \else
            \includegraphics[width=\linewidth]{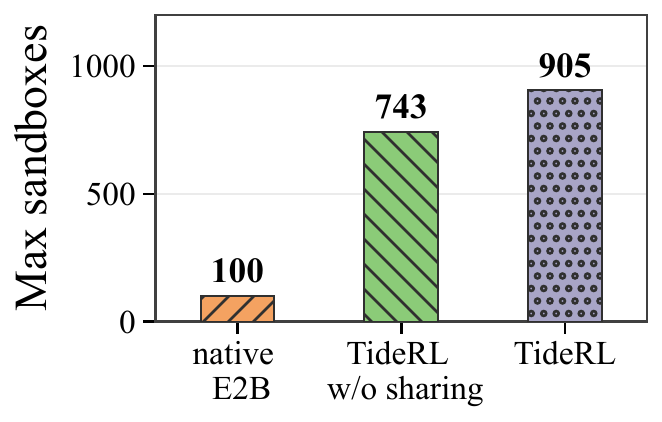}
        \fi
        \vspace{-2\baselineskip}
        \caption{Deployment Density.}
        \label{fig:densitya}
    \end{minipage}
    \hfill
    \begin{minipage}[b]{0.48\linewidth}
        \centering
         \ifarxiv
            \includegraphics[width=\linewidth]{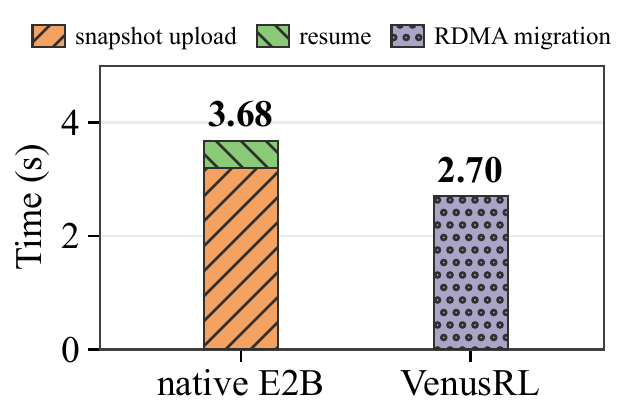}
        \else
            \includegraphics[width=\linewidth]{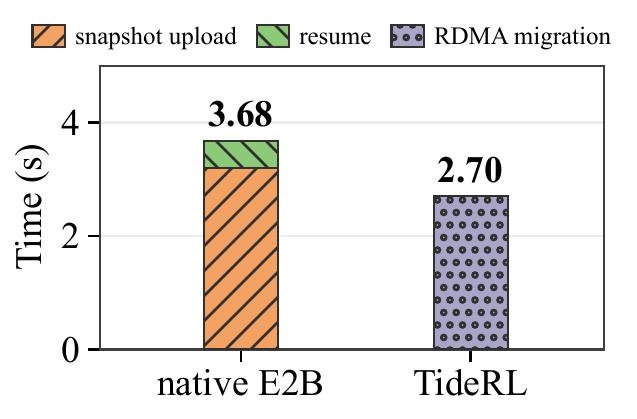}
        \fi
        \vspace{-1.5\baselineskip}
        \caption{Migration Overhead.}
        \label{fig:migration}
    \end{minipage}
    \vspace{-1\baselineskip}
\end{figure}


\noindent\textbf{Ablation study.}
In this experiment, we validate the necessity of the key components of \X's environment resource manager.
We colocate sandboxes cloned from the same environment template on one node with 400GB of host memory, and compare three different configurations. 
Figure~\ref{fig:densitya} shows the maximum number of sandboxes a node can sustain.
Native E2B saturates at 100 since it admits sandboxes by their declared memory upper limit.
Enabling \X's sandbox placement (\S\ref{51}) alone, which admits sandboxes based on their actual physical footprint, lifts the node to 743 sandboxes (643\% gain).
Adding page sharing (\S\ref{52}) further shrinks each sandbox's footprint,
pushing density to 905, 905\% compared with native E2B and 122\% compared with \X without page sharing, indicating the strong scalability of \X.

\noindent\textbf{Sandbox Migration Performance.}
This experiment evaluates \X's RDMA-based sandbox migration mechanism.
We compare \X with the native E2B's recovery path and measure the end-to-end recovery latency of both approaches.
The baseline pauses sandboxes on the source node, uploads saved snapshot and resumes it on a new node.
\X instead pauses the sandbox, migrates its runtime state to the destination node over RDMA, and resumes it there.
As shown in Figure~\ref{fig:migration}, \X recovers a sandbox in 2.70~s, compared with 3.68~s for the native E2B path, a 1.4$\times$ reduction in cross-node recovery latency.

\begin{table}[t]
\centering
\caption{Latency (s) of sandbox lifecycle operations under \X and native E2B.}
\vspace{-1\baselineskip}
\label{tab:overhead}
\footnotesize
\setlength{\tabcolsep}{5pt}
\begin{tabular}{l cc cc cc}
\toprule
 & \multicolumn{2}{c}{128} & \multicolumn{2}{c}{256} & \multicolumn{2}{c}{512} \\
\cmidrule(lr){2-3}\cmidrule(lr){4-5}\cmidrule(lr){6-7}
Operation & E2B & \X & E2B & \X & E2B & \X \\
\midrule
Create & 8.1 & 8.1  & 13.0 & 13.7 & 29.6 & 29.6 \\
Exec   & 3.5 & 4.0  & 4.3  & 4.3  & 3.6  & 3.6  \\
Pause  & 4.9 & 6.8  & 19.2 & 24.8 & 52.3 & 52.6 \\
Resume & 1.7 & 2.1  & 4.9  & 6.2  & 36.1 & 36.2 \\
\bottomrule
\end{tabular}
\vspace{-1\baselineskip}
\end{table}

\noindent\textbf{Overhead of Environment Resource Management.}
To evaluate the overhead introduced by Environment Resource Manager, we compare \X with native E2B under four different primitives (i.e., Create, Exec, Pause, and Resume) with the number of concurrent sandboxes from 128 to 512.
Table ~\ref{tab:overhead} shows the average latency.
Overall, \X introduces only negligible overhead over native E2B.
\vspace{-0.5\baselineskip}
\section{Related Work}
\label{9}

\noindent\textbf{KV Cache Management.}
KV cache management has been extensively studied for efficient LLM serving, primarily focusing on allocation, reuse, and eviction.
PagedAttention ~\cite{kwon2023vllm} manages KV caches with page-like blocks to reduce fragmentation, while vAttention ~\cite{prabhu2025vattention} further decouples virtual and physical KV allocation. 
Other works like Mooncake ~\cite{qin2025mooncake} and HiCache ~\cite{hicache} extend KV storage across GPU memory, host memory, SSDs, or remote storage.
\X's priority-aware KV cache residency is complementary to these techniques and can be integrated with them.

\noindent\textbf{Resource Auto-scaling.}
Auto-scaling is widely used to improve deployment efficiency in cloud systems.
For LLM inference, recent systems~\cite{xiang2025aegaeon,zhao2025seallm,yu2025lambda} optimize resource scaling for large-scale serving workloads.
In RL systems, DistRS~\cite{zhu2026distrs} uses history-based scaling and timeout-aware mechanisms to adjust resources at batch boundaries.
Due to the strong isolation of MicroVM-based environments, \X does not explore dynamic scaling, but we view it as a promising direction for future work.

\vspace{-0.5\baselineskip}
\section{Conclusion}
\label{10}

This paper presents \X, a fully disaggregated agentic RL system with priority scheduling and scalable interaction.
\X designs the priority-aware action scheduler for effective multi-turn rollout scheduling and environment resource manager for scalable and cost-efficient environment interaction.
Implementation and evaluations demonstrate that \X accelerates training by up to 4.24$\times$ and reduces environment costs by nearly 89\%.



\bibliographystyle{ACM-Reference-Format}
\bibliography{reference}
\newpage
\appendix
\section*{Appendix}
\label{sec-appendix}

\section{Analysis on sandbox density bottleneck}
\label{appendix-sandbox-density}

Memory, not CPU or network, is the deployment bottleneck. 
On a representative node in production (128 vCPUs, 512 GiB RAM, 400 GiB for sandboxes; 2 vCPU/4 GB per sandbox), memory caps the node at 100 sandboxes while CPU is only 1.56x oversubscribed. 
Sandbox memory (hugetlb-backed) cannot be swapped or compressed.
Once exhausted, sandbox creation fails, limiting rollout batchsize and staleness, which back-pressures training throughput. 
CPU, by contrast, is a tunable QoS-for-density tradeoff. 
\X's 905-sandbox density implies ~14× oversubscription, viable because agentic sandboxes mostly idle waiting for tools. 
Table ~\ref{tab:overhead} confirms primitive latencies match native E2B. 
Network is likewise a QoS-for-density tradeoff-contention degrades rather than failing sandbox operation, and is rarely observed in production because network bursts seldom time-aligns.

\section{NP-Hardness of Critical-Group identification}
\label{appendix-np-hard}

We consider the decision version of the objective in Section ~\ref{sec_problem_formulation_priority} of the main paper: given unfinished groups, a target $B$, and a deadline $D$, does there exist a schedule that completes at least $B$ groups by $D$?

We use a stateful execution model. 
Initial request placement is unrestricted, and a request may be preempted and resumed on the same worker. 
After a request has performed positive processing on one worker, continuation on another worker requires either state transfer or reconstruction.
Under state transfer, if the request last executes on its source worker at time $s$, its latest state becomes available for continuation on the destination no earlier than $s+\lambda$, where $\lambda>0$ is a handoff-delay parameter of the instance. 
This delay applies only to the migrating request; other requests may execute concurrently.
Alternatively, reconstructing the missing state requires strictly positive additional GPU work. 
Requests do not have free coherent replicas of their execution state.

\noindent\textbf{Theorem.}
Computing an optimal group-completion schedule is NP-hard under this model, even with two identical unit-capacity GPU slots, single-request groups, known processing times, no environment delays, and sufficient KV-cache capacity.

\noindent\textbf{Proof.}
We reduce from \textsc{Partition}.
Given positive integers $a_1,\ldots,a_n$ with
\[
    \sum_{j=1}^{n} a_j = 2T,
\]
construct $n$ groups, each containing one request with processing requirement $p_j=a_j$.
All requests are ready at time zero and initially have no cached execution state. 
Their processing cannot be reduced through cross-request state reuse. 
Set
\[
    B=n, \qquad D=T, \qquad \lambda=T+1.
\]
The two GPU slots can perform at most $2T$ units of GPU work before the deadline, exactly equal to the total processing requirement.

If the \textsc{Partition} instance has a solution, assign the requests corresponding to each subset to one GPU slot.
Each slot receives exactly $T$ units of work. 
Executing its requests sequentially completes all groups by $D=T$, without requiring migration or reconstruction.

Conversely, suppose a feasible schedule completes all $n$ groups by $T$. 
No request can continue on another worker using transferred state: after any source execution ending at $s\geq 0$, destination execution cannot begin before
\[
    s+\lambda = s+T+1 > T.
\]
State reconstruction cannot enable cross-worker continuation either. 
It would add strictly positive GPU work beyond the required $2T$ units, exceeding the total capacity available before the deadline. 
This remains true even if reconstruction overlaps other computation, since it consumes the same aggregate GPU-slot capacity.

Therefore, every request performs all its required processing on the worker where it first executes, although it may be preempted and resumed there.
Let $S$ denote the requests executed on the first slot.
Feasibility requires
\[
    \sum_{j\in S} a_j \leq T,
    \qquad
    \sum_{j\notin S} a_j \leq T.
\]
Since the two sums total $2T$, both equal $T$.
Thus $S$ defines a valid solution to the \textsc{Partition} instance.

The construction has polynomial encoding size, including the handoff delay $\lambda=T+1$.
Hence the deadline-feasibility problem is NP-hard, and so is minimizing the time at which $B$ groups become ready.

\section{Effect of priority scheduling on model learning}
\label{appendix-priority-scheduling-reward}

Though there are concerns that favoring easy-to-complete groups could persistently starve long trajectories, biasing the training data. 
Actually, this does not occur, due to following reasons.

First, priority scheduling does not directly introduce additional off-policy behavior.
In fully asynchronous RL, the off-policy degree of a sample is determined by the mismatch between the behavior policy $\pi_b$ that generates it and the current policy $\pi_\theta$ used for optimization, typically characterized by the importance ratio $r(s,a)=\frac{\pi_\theta(a\mid s)}{\pi_b(a\mid s)}$.
Scheduling only changes the execution and completion order of outstanding trajectories; it does not modify either policy itself. 
Its only possible effect on $r(s,a)$ is indirect, through the number of weight updates that occur before a trajectory is consumed, i.e., its policy staleness.
Therefore, priority scheduling may change the realized staleness of individual trajectories within this window, but does not introduce a new source of off-policy behavior beyond the staleness already present in fully asynchronous training.

Second, long low-priority trajectories are not continuously blocked. 
Fully asynchronous RL does not allow rollout generation to advance arbitrarily far ahead of training.
With global batch size $B$ and maximum staleness $\eta$, only $O((\eta+1)B)$ outstanding trajectories can remain within the admissible asynchronous window.
Length-based priority therefore only reorders a finite competing set.
Short trajectories may temporarily overtake a long trajectory, but they leave the active set once completed, while the staleness controller prevents an unbounded stream of new trajectories from continuously entering ahead of it. 
Additionally, the staleness ratio is deliberately kept low to preserve algorithmic stability. 
Prior works like Asynchronous RLHF ~\cite{noukhovitch2025asynchronous} and StreamRL ~\cite{zhong2025streamrl} recommend that accumulated stale samples should not exceed one global batch size. 
As the step progresses, the higher-priority competing set shrinks, allowing long trajectories to continue generation rather than being repeatedly demoted.
Thus, the stale long trajectory naturally becomes high-priority samples when step progresses.

Third, \X adds a hard safeguard for the tail case. 
Once a trajectory reaches its last admissible weight version, \X immediately promotes it to the highest priority until completion. 
At this point, ordinary length-based priority no longer applies, preventing newer short trajectories from further delaying it. 
Thus, bounded staleness already limits the opportunity for persistent deprioritization, while staleness-boundary promotion provides an explicit safety guard against priority-induced starvation.
Moreover, \X is fully compatible with off-policy correction methods such as TIS ~\cite{yao2025rollout} and IcePop ~\cite{zhaosmall}, further preserving stability.

\section{Impact of OOM Recovered Behaviour}
\label{appendix-OOM}
\X does not rely on a single mechanism to prevent OOM. 
Admission Threshold Adaption (Algorithm~\ref{alg}) serves as the primary safeguard: a sandbox creation request is admitted only when the projected memory footprint of resident sandboxes remains below a workload-specific safety threshold. 
The reserved headroom is intended to absorb memory growth that occurs after admission and between consecutive control intervals.

In practice, such growth is both sparse and small. 
In our production trace, fewer than $0.5\%$ of the sandboxes on a node exhibited concurrent memory bursts within a control interval, and each burst increased memory usage by at most roughly $100$ MiB. 
These events were primarily caused by on-demand package installation triggered by agent tool calls and were therefore short-lived rather than sustained. 
Consequently, the aggregate unaccounted memory growth within a control interval is usually well within the headroom reserved by the safety threshold. 
With the threshold empirically tuned for each workload, the observed OOM frequency is therefore near zero. 

Additionally, rather than provisioning substantially more memory for these rare cases, \X treats OOM as a recoverable event. 
When it occurs, the affected sandbox is recovered on another node by transferring its full memory state over P2P RDMA and resuming execution there. 
This recovery takes approximately $2.7$ s (Figure~\ref{fig:migration}), substantially less than the duration of a single RL step, making its overhead negligible at the RL timescale. 
We also further manually inspected recovered sandboxes and verified that both filesystem contents and running process state were preserved across recovery.

\section{Robustness of the predictor}
\label{appendix-predictor-robustness}

\X uses predicted remaining generation lengths to separate long and short samples and prioritize them accordingly.
Thus, the scheduler primarily relies on the relative ordering of predictions rather than accurate absolute token counts.
Prediction errors are therefore tolerable as long as they preserve the ordering relevant to scheduling.
The predictor conditions on both the problem specification and the accumulated trajectory history: the former provides signals about task difficulty, while the latter---including intermediate reasoning, tool calls, and environment observations---reveals the sample's current progress and remaining work.
As execution proceeds, the predictor updates its estimate at turn boundaries, allowing scheduling decisions to track the evolving trajectory state.
Policy updates during RL may shift absolute generation lengths, but they do not necessarily change the relative ordering across samples.
Consequently, a lightweight predictor fine-tuned before training can remain effective as the rollout policy evolves.
The Spearman correlation of $0.98$ reported in the main paper further confirms that the predictor accurately recovers the ordering required for priority scheduling.
Additionally, the predictor can also be updated at training checkpoints using recent trajectories to mitigate policy drift, with fine-tuning typically converging within only minutes.

We train the predictor using trajectories collected at the same sampling temperature as the target rollout workload.
For each temperature, we collect a separate set of trajectories and construct the corresponding \texttt{(messages, length)} pairs for fine-tuning.
This matching avoids distribution shift between predictor training and rollout-time sampling behavior.
Supporting a different rollout temperature therefore only requires fine-tuning on trajectories collected under that temperature; the prediction mechanism itself is not tied to any particular setting.
We report results at temperature $0.7$ since this is the SWE-agent configuration used throughout our evaluation, rather than a limitation of the predictor.

\end{document}